\documentstyle[emulateapj,psfig]{article}

\received{}
\accepted{}
\journalid{}{}
\articleid{}{}
\slugcomment{Accepted for publication in ApJ: scheduled for the March
20, 1999 issue, Vol. 514.}
\lefthead{}
\righthead{}

\def\p{\prime}

\begin{document}

\title{  
A Flux-limited Sample of Bright Clusters of Galaxies from the Southern Part of
the ROSAT All-Sky Survey: the Catalog and the LogN-LogS.
}

\vskip 30pt
\author{S. De Grandi\altaffilmark{1,2}, 
        H. B\"ohringer\altaffilmark{1}, 
        L. Guzzo\altaffilmark{2}, 
        S. Molendi\altaffilmark{3}, 
        G. Chincarini\altaffilmark{2,4}, 
        C. Collins\altaffilmark{5},
        R. Cruddace\altaffilmark{6},
        D. Neumann\altaffilmark{7},
        S. Schindler\altaffilmark{1,5}, 
        P. Schuecker\altaffilmark{1}, and
        W. Voges\altaffilmark{1}
}

\altaffiltext{1}{Max-Planck-Institut f{\"u}r extraterrestrische Physik, 
Giessenbachstra$\ss$e 1, 85740 Garching bei M\"unchen, Germany.}
\altaffiltext{2}{Osservatorio Astronomico di Brera, via Bianchi 46,
22055 Merate (LC), Italy.}
\altaffiltext{3}{Istituto di Fisica Cosmica, CNR, via Bassini 15, 
20133 Milano, Italy.}
\altaffiltext{4}{Universit\`a di Milano, via Celoria 16, 
20133 Milano, Italy.}
\altaffiltext{5}{Astrophysics Research Institute, Liverpool John Moores 
University, Byrom-Street, Liverpool L3 5AF, United Kingdom.}
\altaffiltext{6}{E. O. Hulburt Center for Space Research, Naval Research 
Laboratory, Code 7620, 4555 Overlook Ave., Washington, DC 29375.}
\altaffiltext{7}{CEA/Saclay, Service d'Astrophysique, L'Orme des Merisiers 
Bat.709, 91191 Gif-sur-Yvette, France.}

\submitted{Accepted for publication in ApJ: scheduled for the March
20, 1999 issue, Vol. 514.}
\righthead{DE GRANDI ET AL.}
\lefthead{CATALOG OF SOUTHERN ROSAT ALL-SKY SURVEY CLUSTERS}

\begin{abstract}
We describe the selection of an X-ray flux-limited sample of bright 
clusters of galaxies in the southern hemisphere, based on the first 
analysis of the ROSAT All-Sky Survey data (RASS1).  
The sample is constructed starting from an identification of candidate
clusters in RASS1, and their X-ray fluxes are re-measured using the Steepness 
Ratio Technique.  
This method is better suited than the RASS1 standard algorithm for measuring 
flux from extended sources.
The final sample is count-rate-limited in the ROSAT hard band (0.5-2.0 keV), 
so that due to the distribution of $N_H$, its effective flux limit varies 
between $\sim 3.$ and $4. \times 10^{-12}$ erg cm$^{-2}$ s$^{-1}$ 
over the selected area.  
This covers the $\delta<2.5^o$ part of the south Galactic cap region
($b_{II}<-20^o$) -- with the exclusion of patches of low RASS1 exposure 
time and of the Magellanic Clouds area -- for a total of 8235 deg$^2$.  
130 candidate sources fulfill our selection criteria for {\it bonafide} 
clusters of galaxies in this area.  
101 of these are Abell/ACO clusters, while 29 do not have a counterpart in 
these catalogs.  
126 ($97\%$) clusters have a redshift and for these we compute an X-ray 
luminosity.
$20\%$ of the cluster redshifts come from new observations, as part of the
ESO Key Program REFLEX Cluster Survey that is under completion.
Considering the intrinsic biases and incompletenesses introduced by
the flux selection and source identification processes, we estimate
the overall completeness to be better than $90\%$.
The observed number count distribution, LogN-LogS, is well fitted by a power 
law with slope $\alpha = 1.34\pm 0.15$ and normalization $A=11.87\pm 1.04$ 
sr$^{-1}$ (10$^{-11}$ erg cm$^{-2}$ s$^{-1})^{\alpha}$, in good agreement 
with other measurements.
\end{abstract}


\section{Introduction}
Clusters of galaxies are astrophysical objects carrying fundamental 
information about the characteristics of the Universe. 
They are the largest gravitationally bound units, and the 
time scales involved in their dynamical evolution
are comparable to the age of the Universe.  
Hence clusters are cosmologically young structures, in which evidence of 
the initial conditions has not yet been totally removed by dissipative effects.
On the other hand their dynamical state is approaching a characteristic 
equilibrium configuration that allows a coherent modeling (e.g. King 1966).
Because of these particular properties, the characteristics of the
population of galaxy clusters are strongly related to the cosmological 
parameters.
Both theory and simulations have shown that cluster morphologies,
comoving densities, and clustering properties provide information on
the density parameter, $\Omega_0$, and on the shape and normalization of the 
primordial power spectrum
(e.g., Frenk et al. 1990; Bahcall \& Cen 1993; White et al. 1993;
Kitayama \& Suto 1997; Evrard 1997).

To be able to measure such ensemble--averaged quantities, one needs
to select large, statistically complete samples, which have to be 
representative of clusters of galaxies as a class.

In parallel with deep cluster surveys (e.g. Rosati et al. 1997; Collins
et al. 1997; Jones et al. 1998), it is
particularly important to accurately define the properties
of the cluster population at low redshifts, to provide 
the reference frame for quantifying cluster evolution.
For example, an accurate estimate of the luminosity function, which 
is an essential
part of this statistical analysis, requires the detection of clusters over
a fairly large volume of space to reach an accurate evaluation of the bright
end of the distribution.
On the other hand, a sample of clusters of galaxies covering a large 
volume of the local Universe, would allow the
study of clustering over scales $>> 100$ h$^{-1}$ Mpc, i.e. around the 
turnover of the power spectrum.  
On these scales, these measures would 
overlap the most recent and future microwave background anisotropy
experiments (e.g. Gawiser \& Silk 1998), thus providing a direct
comparison of the clustering in the light and in the mass. 

In recent years it has become ever more evident that the 
selection of large samples of clusters is best done in 
the soft X-ray band (i.e. energies in the range from $\sim$ 1 to 10 keV), where
clusters are prominent among the various classes
of extragalactic objects by virtue of their extremely high X-ray luminosities,
$\sim 10^{43}-6\times 10^{45}$ erg s$^{-1}$ (e.g. B\"ohringer 1995). 
The X-ray emission originates in the thin hot gas, contained by the 
deep cluster potential well, in a state approaching the hydrostatic
equilibrium (e.g. Sarazin 1988).
The predominant emission mechanism at these high energies in clusters
is thermal bremsstrahlung, and as 
the hot gas is distributed throughout the potential well, clusters of galaxies
observed in the X-ray band appear, unlike at optical wavelengths, as
single diffuse entities.
Moreover, as the X-ray intensity scales quadratically with the gas density,
whereas the integrated optical luminosity is 
only linearly correlated with the galaxy density,
observations in the X-ray band are less subject to the projection 
effect problems which affect optically selected catalogs of galaxy clusters.

A unique opportunity to construct large cluster samples selected in the
X-ray band has been provided by the ROSAT mission (Tr\"umper 1993).
This satellite conducted, in the second half of 1990, an all-sky survey 
in  soft X-rays (0.1-2.4 keV) with the Position Sensitive Proportional
Counter (PSPC; Briel \& Pfeffermann 1986) as focal plane detector.
Due to the improved spatial resolution, the higher sensitivity, and the 
smaller intrinsic background with respect to previous 
X-ray satellites, ROSAT data are especially attractive for the 
study of clusters of galaxies.
Furthermore, the ROSAT All-Sky Survey (RASS) is the first conducted
with an imaging X-ray telescope and is uniquely suited to obtaining
complete X-ray cluster samples over large areas of the sky.
Several approaches to the problem of selecting statistical samples of galaxy
clusters from the RASS1 database have been described (e.g.
Burg et al. 1992; Briel \& Henry 1993; Romer et al. 1994;
Ebeling et al. 1996, 1997).

In 1992 an ESO Key Program (hereafter ESOKP; B\"ohringer 1994; 
Guzzo et al. 1995) was started with the aim of constructing in the 
southern sky the largest flux limited sample of clusters of galaxies 
using the RASS.
Due to the huge number of X-ray sources detected, more than 50000 objects,
it was not feasible to start an observing campaign with
the aim of optically identifying all the RASS sources in the southern
hemisphere.
However, it is generally assumed that mass fluctuations on scales of a few to
tens of Megaparsecs lead to  the formation of clusters of galaxies visible
optically and in the X-ray band. Therefore the approach followed by the
ESOKP collaboration is to search for correlations between X-ray sources
with regions of galaxy overdensity.
The resulting list is then correlated with catalogs such as the NASA
Extragalactic Data Base (NED) and SIMBAD, in order to remove chance 
correlations with stars and AGNs.
The final step is to observe the refined list spectroscopically, in order to
derive the cluster redshift, if not already known, and to remove some
remaining sources which are not clusters.

As discussed in De Grandi et al. (1997, Paper I hereafter), however, the
standard analysis (see $\S$ 2.1) performed on the RASS data is not fully 
appropriate for characterizing extended sources. 
In particular, fluxes are systematically underestimated.
These limitations have prompted the development of alternative techniques
such as the Steepness Ratio Technique (SRT), discussed in Paper I, and
the Voronoi Tesselation and Percolation analysis (Ebeling et al. 1996).  
Here, we present the application of the SRT to an initial set of candidate 
clusters from the ESOKP, that leads to the construction of a complete flux 
limited sample of bright clusters of galaxies.  
The sample is limited to the southern Galactic cap region 
($b_{II} < -20^o$, $\delta < 2.5^o$), and its X-ray and optical completeness 
are investigated in detail.  
In addition, we shall also compute and discuss the LogN-LogS distribution.

A preliminary version of the sample presented here, was previously used 
to obtain an estimate of the cluster X-ray luminosity function 
(XLF, De Grandi 1996b).
An updated estimate of the XLF from the present catalog is the subject 
of a parallel paper (De Grandi et al. 1999).  
These results will be extended by the future developments of the 
ongoing ESOKP collaboration (now known as the REFLEX Cluster Survey,
see B\"ohringer et al. 1998).
 
The paper is organized as follows.
In  $\S$ 2 we describe briefly the RASS1 data used to derive the X-ray 
source properties and the various algorithms applied to these data. 
In $\S$ 3 we describe the initial procedures used by the ESOKP collaboration 
to define a sample of clusters in the southern hemisphere using 
the RASS1 data, which we call here the RASS1 Candidate Sample. 
The procedure for the selection and definition of the sample of bright 
clusters, which we shall call the RASS1 Bright Sample, is presented 
and discussed in detail in $\S$ 4. 
In $\S$ 5 we compute the number counts, or LogN-LogS distribution,
of this sample, and compare it with previous results.
In $\S$ 6 we investigate the potential biases that could affect the 
RASS1 Bright Sample.
In $\S$ 7 we summarize our main results and conclusions.

\section{Analysis of ROSAT All-Sky Survey Sources}

\subsection{SASS1 analysis of strip data}
The first analysis of the ROSAT All-Sky Survey sorted the data into 90 
great-circle strips on the sky, each $2^o$ wide and passing through the
ecliptic poles.
These strips were processed one by one by using the Standard Analysis 
Software System (SASS1; e.g., Voges 1992) developed at MPE  
(Germany).
The detection process used photons in the broad PSPC energy band 0.1-2.4 keV.
Each data strip was analyzed using a combination of source detection 
algorithms, including two {\it sliding window} techniques (the first 
using a local background determination, the latter a global background 
map) and a Maximum Likelihood (ML) method (Cruddace et al. 1988).
SASS1 produced an all-sky source catalog of $\sim 50.000$ objects
(with existence likelihood larger than 10)
containing information about source properties, such as
detection significance, count rate, position, hardness ratio and extent.
As explained in $\S$ 3 the first candidate sample for the ESOKP,
on which the present work is based, was selected from this catalog.

\subsection{Analysis of merged data}
As mentioned in the introduction, our first goal was to re-estimate 
using the SRT approach the X-ray fluxes for all candidate clusters, 
selected from the SASS1 source list.
After the end of the survey, it was possible to obtain from the ROSAT
team $2^o\times 2^o$ sky fields, centered on the SASS1 ML positions, 
containing the photons accumulated from all the overlapping strips, 
known as {\it merged data}.
We collected the merged data fields for all our cluster candidates and 
first analyzed them using the standard detection procedure, as
implemented in the EXSAS package (Zimmermann et al. 1997).  
The first aim of this was to re-estimate the source positions with the ML 
algorithm with improved accuracy.
The standard analysis on the merged data was performed in the three ROSAT 
PSPC energy bands: the broad band (0.1-2.4 keV), the soft band (0.1-0.4 keV) 
and the hard band (0.5-2.0 keV).
A detailed description of this analysis can be found in Paper I.

We proceeded then by applying the SRT to the merged data using the new ML
position.
We recall here that SRT uses the convolution between the real RASS
point-spread function (G. Hasinger 1995, private communication) and a 
$\beta$-model of the cluster emission profile (with parameter $\beta$ 
fixed to 2/3), $\tilde I(r)$, to derive for each cluster the core radius 
and the total count rate.
We compute for each source the observed steepness ratio, SR$_{obs}$, 
that is the ratio between the source counts within an annulus bounded by 
the two radii of 3 and 5 arcmin and those within a 3 arcmin
radius circle.  The source core radius is derived by comparing 
SR$_{obs}$ with the theoretical SR = $\int^{5^\p}_{3^\p} 2\pi r \tilde 
I(r) dr/ \int^{3^\p}_{0^\p} 2\pi r \tilde I(r) dr$, 
as a function of the core radius (Figure 5 in Paper I).
The total source counts are computed by correcting the measured counts
within 5 arcmin radius, by the fraction falling outside this aperture, by means
of an SR$_{obs}$ dependent correction factor $F$, shown in Figure 8 of Paper I.
SRT also evaluates the probability for each source to be pointlike, which
is computed without using a specific model for the source emission profile.
The SRT analysis was performed in all three ROSAT PSPC energy bands.

\section{The RASS1 Candidate Sample}
As soon as the ROSAT All-Sky Survey went through its first processing
(SASS1), an automatic identification program
for galaxy clusters in the Southern sky was set-up as a collaboration
between MPE and ROE/NRL.  
The broad aim of this work was to identify all SASS1 X-ray sources that 
could possibly be associated with a cluster of galaxies, known or unknown 
from available optical cluster catalogs.   
Given the detection limit of the RASS, the redshift distribution of the 
identified clusters was expected to peak between 0.1 and 0.2 in 
redshift, so that the bulk of the rich cluster population was clearly 
detectable on the ESO/SRC survey plates.  

The main identification method was therefore to look for overdensities of 
galaxies in the ROE/NRL object catalog, produced by digitizing the 
IIIa-J plates with the COSMOS machine in Edinburgh (Yentis et al. 1992), 
around each SASS1 source above a given SASS1 count rate (see below).  
An excess probability could thus be defined by comparison with 
counts at random positions and well characterized thresholds
in completeness and contamination could be defined.
Relatively low search thresholds in contamination were used to avoid 
discarding genuine X-ray clusters, leading to the inclusion in the candidate 
list also of spurious objects that had to be removed in a later step of the 
work.
As a further complement to this, the SASS1 source list was also correlated 
with the Abell/ACO catalog of clusters of galaxies (Abell et al. 1989) and 
with a catalog of automatic clusters independently constructed from the 
COSMOS galaxy database.   

The third method was to include also all the sources flagged 
by SASS1 as having an extent radius $>25$ arcsec and an
extent likelihood $>7$.
This is the only method which is based on the X-ray properties alone.  
Its main drawback is that of not being particularly 
robust, so that some truly extended sources are not recognized as such.  
In addition, the RASS data are not optimal in general for recognizing 
extended sources, due to the low photon statistics that couples with
the limited resolution of the RASS.  
Despite its limitations, however, this technique complements the 
optically--based methods, and (\S 6.2), provides a useful way to roughly 
estimate the completeness of the identification process.
Finally, the three cluster candidate lists were merged and multiple X-ray 
detections of the same object were removed, for a total of $\sim 1000$ 
objects.
In this paper we consider only southern sky candidates at high Galactic
latitudes, $\delta < 2.5^o$ and $b_{II} < -20^o$, leading to a sample of 679.
This list will be referred to as the RASS1 Candidate Sample.

As mentioned above, prior to the identification procedure, the initial 
sample of sources from RASS1 was thresholded to a count rate limit of
0.055 cts/s. 
Due to a problem in the early processing of the data, however, the count
rate in a few strips had a systematic shift, so that in the end they
were thresholded to a higher limit, 0.08 cts/s. 
This problem, discovered later in the course of our analysis, affected 
$17\%$ of the total sample.
We shall show in the following how this residual incompleteness has been
taken into account.

\section{Selection of the RASS1 Bright Sample}
In this section we describe how we proceeded in order to select 
a flux limited sample of clusters of galaxies from the 
RASS1 Candidate Sample.
To this end we must first consider the possible sources of 
incompleteness that could affect the sample.

\subsection{Exposure Times and Sky Area Selections}
A first source of incompleteness derives from the differences in exposure
times over the different strips, which in the southern sky may vary 
between 0 and $\sim$ 800 seconds, with a peak at $\sim 400$ seconds.
In Figure 1 we plot the SASS1 broad band count rates versus the SASS1 
exposure times for the RASS1 Candidate Sample. 
We separate the pointlike sources (asterisks) from the extended ones 
(open circles) by using the SRT probability of extension: we define as 
extended any source with a probability of being pointlike $< 1$\%.
The solid and dotted lines drawn in Figure 1, correspond to the count rate 
limits of 0.055 cts/s and 0.08 cts/s discussed at the end of $\S$ 3, 
respectively.  
Note how  for both count rate limits, SASS1
starts to fail in detecting sources when the exposure time becomes smaller 
than $\sim 100-120$ s. 
Therefore, we consider only regions with exposures larger than 150 seconds, 
in order to avoid regions of the sky where objects could have been 
missed because of the low sensitivity of the survey.

To avoid incompleteness problems related to the difficulty in identifying 
clusters inside optically crowded fields, we also excluded the sky areas
of the Galactic plane and the Magellanic Clouds.
The Galactic plane region was in fact already excluded in the early
selection of the RASS1 Candidate Sample to avoid regions of high $N_H$ values,
by selecting only sources with $b_{II} < - 20^o$.
To establish the size of the area to be excluded in the case of the 
Magellanic Clouds, we considered both the optical and the X-ray Cloud
emission (e.g. Snowden \& Petre 1994). 
We opted for a conservative choice rejecting an area 
slightly larger than that covered by both the emissions. 
The lower left and upper right corners of the excluded areas in equatorial 
coordinates (J2000.0) are
$\alpha_{ll} = 93.135^o, \delta_{ll} = -77.5167^o$  and 
$\alpha_{ur} = 60.1783^o, \delta_{ur} = -62.3611^o$ for the LMC, and 
$\alpha_{ll} = 22.663^o, \delta_{ll} = -77.243^o$  and 
$\alpha_{ur} = 353.223^o, \delta_{ur} = -67.224^o$ for the SMC, 
respectively.

After excluding these areas and setting a threshold for the exposure time 
we are left with the geometric area of 8235 deg$^2$ (i.e., 2.5 sr), which 
is shown in Figure 2.
This is about a fifth of the whole sky, or a third of the sky available 
at $|b_{II}|>20^o$. 
The number of cluster candidates in this area is 540.

\subsection{Count Rate Selection}
In $\S$ 3 we discussed how the RASS1 Candidate Sample was selected, starting
from the list of X-ray sources detected in the survey by SASS1.
However this initial selection procedure was lacking in two respects.
First, a flux threshold should be set using count rates in the ROSAT
hard band (0.5-2.0 keV), which is best suited for the analysis of 
hard sources such as clusters, 
whereas initially thresholds were set using SASS1 count rates in the
broad band (0.1-2.4 keV).
Second, the SASS1 algorithm, designed for speed, was rather imprecise in 
estimating the flux and angular extent of clusters (Paper I; Ebeling 
et al. 1996), a problem the SRT algorithm was designed to correct.
In this section we select bright clusters from the RASS1 Candidates Sample 
by means of a cut in SRT count rate computed in the hard band. 
We show that a limiting SRT count rate of 0.25 cts/s leads to a sample
characterized by a high degree of completeness.

\subsubsection{Study of a Control Sample}
In order to understand the effects of changing the energy band and introducing
a further count rate selection using SRT results we investigate
the behavior of a control sample of sources, obtained from the
{\it Einstein} Extended Medium Sensitivity Survey (EMSS; Gioia et al. 1990; Maccacaro et al. 1994) and reobserved in the RASS.
No cut in SASS1 count rates has been applied to this dataset.
Out of the 835 EMSS objects we selected two samples, the first comprising
pointlike sources, i.e. objects classified as AGNs, BL Lac and stars 
(Maccacaro et al. 1994), and the second potentially extended
sources, i.e. objects classified as galaxies or clusters of galaxies.
For both samples we include only objects detected in the RASS1 merged data
by the ML algorithm (see $\S$ 2.2) in the broad band, 
with an ML existence likelihood larger than 12.
These selections lead to well defined EMSS control samples of 108 
pointlike and 50 potentially extended objects.

In Figure 3 we show the comparison between the SRT hard band count rates
and the SASS1 broad band count rates for the EMSS pointlike (asterisk) and
potentially extended (open circles) sources.
The observed distribution 
allows us to estimate how many sources would be lost in passing from
one X-ray analysis system to the other.
The sources lost due to the SASS1 count rate cut applied prior to the SRT
analysis should fall in the top left quadrant delimited by the
dotted vertical and horizontal lines representing the cuts of 0.055 cts/s 
and 0.25 cts/s, respectively.
Since no pointlike sources are present in the top left quadrant, 
we deduce that the degree of completeness of a sample of pointlike sources 
with a threshold in the SRT 
hard band count rate of 0.25 cts/s, previously cut with a SASS1 broad band 
count rate of 0.055 cts/s, is extremely high. 
Two ``extended'' objects fall in the top left quadrant, indicating 
that the completeness of the corresponding sample of extended sources, 
although still quite high, is not $100\%$.
The difference between the distribution of pointlike and extended sources
in the count rate plane is produced by the strong underestimation of the 
count rate by SASS1 for extended sources (see $\S$ 4.2).

\subsubsection{Study of the RASS1 Candidate Sample}
In the light of the results described above we examine now our list of 
cluster candidates. 
As we mentioned in $\S$ 3, this is divided into two distinct subsamples 
characterized by 
two SASS1 broad band count rate limits (0.055 and 0.08 cts/s). 
We call the first SUB1 and the second SUB2.
SUB1 is not only deeper but also more populated ($83\%$ of the total 
RASS1 Candidate Sample).
In Figure 4 we plot for SUB1 the SRT hard band against the SASS1 
broad band count rates. 
The horizontal and vertical dashed lines correspond to the count rate
limits of 0.25 and 0.055 cts/s, respectively.
The sources we miss when cutting the SUB1 sample at an SRT hard band
count rate of 0.25 cts/s, because of the prior SASS1 cut, are those
which would populate the top left quadrant.
We have used the observed distribution of points in Figure 4 to
estimate the number of missed sources.
After selecting the SUB1 sources with $cr_{SRT}>0.25$ cts/s (133 sources),
we have plotted the distribution of their SASS1 count rates.
This is shown in logarithmic bins in Figure 5.
The histogram has a maximum for $cr_{SASS1}\sim 0.25$ cts/s, and clearly, the 
decrease below this value is produced by the applied SRT cut.
The number of missed sources in the top left quadrant of Figure 4 can then
be estimated extrapolating the distribution below the $cr_{SASS1}$ cutoff.
A linear extrapolation, using the four bins nearest to the SASS1 limit, 
gives a number of missing sources equal to $\sim 1.5^{+5.4}_{-1.5}$ 
In order to be conservative  we will take the upper bound of this result, and
assume that 6.9 sources are lost.

We now wish to estimate the sources we miss when cutting the SUB2 sample 
at SRT count rate of 0.25 cts/s, because of the prior SASS1 cut.
The number of SUB2 sources above $cr_{SRT} = 0.25$ cts/s is 31. 
The number of SUB2 sources which should be falling within the SASS1 count
rate range $0.055-0.08$ cts/s can be directly estimated from the observed
distribution of SUB1 sources: 8 objects of SUB1 fall within the SASS1 range
$0.055-0.08$ cts/s, corresponding to $\sim$ 1.9 objects in SUB2. 
The number of SUB2 sources which should be falling below $cr_{SASS1} = 0.055$ 
cts/s (i.e., $\sim$ 1.3 objects) can then be derived using that estimated for 
SUB1.
Adding all these contributions, we estimate that for the whole RASS1 
Candidate Sample, the expected missing sources amount to $\sim 8.6$ (i.e.
$\sim 5\%$ of the total sample).
We therefore expect the sample to be $\sim 95\%$ complete as far as the X-ray 
selection is concerned.
Such a high degree of completeness has been achieved at
the price of reducing drastically the number of sources:
the RASS1 Candidate Sample contained 679 objects, while 
this sample contains now 164 candidates.

\subsection{Definition of the RASS1 Bright Sample}
As a result of the selections described in the previous subsections we
obtain a list of 164 cluster candidates.
In order to assign a reliable classification to these objects we have 
collected the following information: 
1. images from the Southern Digitized Sky Survey (SDSS); 
2. overlays of X-ray contours on optical object distributions from the 
COSMOS catalog;
3. information from the NED and SIMBAD databases;
4. X-ray properties, such as source extent, probability of
extent, X-ray flux and luminosity from RASS1 data and ROSAT pointed data
when available;
5. optical CCD images and spectra from the ESOKP.
Redshifts were obtained from the literature and from our new ESOKP 
observations.

The analysis of this information allowed us to divide the cluster
candidates into 4 groups:

\begin{list}
{(A)}{} 
\item confirmed clusters of galaxies, i.e. objects for which the 
inspection of optical images and spectra allow a certain identification
of the X-ray source with a cluster;  
\end{list}
\begin{list}{(B)}{} 
\item likely clusters, i.e. objects with X-ray and optical properties 
consistent with clusters, lacking firm spectroscopic confirmation;
\end{list}
\begin{list}{(C)}{} 
\item uncertain identifications, i.e. objects for which either
we have insufficient information to discriminate between two possible
classifications or we have no information at all;
\end{list}
\begin{list}{(D)}{}
\item objects which are certainly not galaxy clusters (27 obj.).
\end{list}

This classification into the four classes was done independently 
three times to reduce subjective definitions, and the agreement was excellent.

The rather high degree of contamination was expected
as a result of the pre-selection methods we have applied in constructing the
RASS1 Candidate Sample. 

In paper I we pointed out that for very extended sources the source profile 
appears to be almost flat between 0 and 5 arcmin and the observed 
steepness ratio 
approaches its maximum value (i.e. the ratio between the areas of the 
annulus bounded by 3 and 5 arcmin and of the circle of 3 arcmin radius).
This large observed steepness ratio leads to diverging values of the 
core radius, the total source counts and the associated uncertainties.
To overcome this effect we impose that the physical core radius (in kpc) of a 
cluster candidate cannot be larger than a certain upper limit.
If the physical core radius found by SRT is larger than the limit 
we compute a new angular core radius from the limiting core radius using 
the redshift of the source, derive the corresponding steepness ratio 
(see Figure 5 in Paper I),
and finally, with the new steepness ratio and the curve shown in Figure
7 in Paper I, we compute the revised count rate of the source.
The average core radius of rich clusters is about 250 kpc (Bahcall 1975), while
Jones and Forman (1984) found that out of 38 clusters $80\%$ have a core 
radius smaller than 300 kpc and only $20\%$ have a core radius in the range
300 and 800 kpc.
Various tests with different values of the physical core radius show
that the best compromise is achieved with a value of 400 kpc
for the upper limit of the core radius.
21 sources belonging to classes A, B and C have a modified SRT count rate, 
but for 14 only the revised count rate differs by more than $10\%$ from its 
original value. 
Moreover, 7 sources have a revised SRT count rate that falls below our 
threshold of 0.25 cts/s and therefore leave the sample.

We define as the RASS1 Bright Sample the sum of sources belonging
to the A (119), B (6) and C (5) groups, using the results of the revised SRT.
The final sample contains 130 sources.
A schematic representation of the selections leading to the construction 
of the RASS1 Bright Sample from the RASS1 Candidate Sample is given 
in Figure 6.

\subsection{The Catalog of X-ray sources}
The sources of the RASS1 Bright Sample are presented in Table 1.
Columns list the observed and derived parameters for each source as follows:

{\it Column (1). ---} Sequence number of the source in the catalog.

{\it Column (2). ---} Position: right ascension (hh mm ss.s, first line) and
declination (dd mm ss.s, second line) as derived by the ML algorithm when 
analyzing the RASS1 merged data (J2000.0 coordinates).

{\it Column (3). ---} Column density of Galactic Hydrogen from Dickey
\& Lockman (1990) in units of $10^{20}$ atoms cm$^{-2}$.

{\it Column (4). ---} Vignetting corrected RASS1 exposure time computed
from merged data in seconds.

{\it Column (5). ---} Source count rate (first line) computed in a circle of 
5 arcmin radius from the source position in the PHA channels 
from 52 to 201 (corresponding to the 0.5-2.0 keV energy band), and 1-$\sigma$ 
errors (second line) from photon-counting statistics.

{\it Column (6). ---} Total SRT source count rate (first line) in the 0.5-2.0 
keV band and associated uncertainties (second line).

{\it Column (7). ---} Model independent probability of the source to be 
pointlike. 
Sources with probability smaller than $1\%$ are considered extended.

{\it Column (8). ---} Unabsorbed X-ray flux (first line) computed in the 
0.5-2.0 keV band in units of $10^{-11}$ erg cm$^{-2}$ s$^{-1}$, and 
associated symmetrized 1-$\sigma$ uncertainties (second line).

{\it Column (9). ---} X-ray luminosity (first line) computed in the 
0.5-2.0 keV band in units of $10^{44}$ erg s$^{-1}$, and associated 
1-$\sigma$ uncertainties (second line). 
The luminosity has been computed in the rest frame of the source by assuming 
a power law spectrum with energy index 0.4, $H_0=50$ km s$^{-1}$ Mpc$^{-1}$
and $q_0=0.5$. 

{\it Column (10). ---} Optical identification. Name of the source 
(first line), if already known, and proposed classification (second line)
according to the description given in $\S 4.3$.

{\it Column (11). ---} Source redshift (first line) and associated reference 
(second line). 

{\it Column (12). ---} Comments: contains miscellaneous information on the
source. 

In Table 2 we list the sources belonging to class D (i.e. non-cluster
objects discarded from the RASS1 Bright Sample).
The contents of Table 2 are:

{\it Column (1). ---} Optical name of the source if already known.

{\it Column (2). ---} Right ascension (J2000.0 coordinates) of the X-ray
source from RASS1 merged data.

{\it Column (3). ---} Declination (J2000.0 coordinates) of the X-ray source 
from RASS1 merged data.

{\it Column (4). ---} Unabsorbed X-ray flux computed in the 0.5-2.0 keV band
with SRT, in units of 10$^{-11}$ ergs cm$^{-2}$ s$^{-1}$.

{\it Column (5). ---} Uncertainties associated to the X-ray flux in the 
same units and energy band as in column (4).

{\it Column (6). ---} Source type. STA = star, GAL = ``normal'' galaxy, 
GC = globular cluster, AGN = Active Galactic Nucleus (Quasar, Seyfert 
galaxy or BL Lac object).

\subsection{Extent of Clusters in the Catalog}
For each cluster of the RASS1 Bright Sample we have computed the
probability to be a pointlike source following the SRT method described
in Paper I.
We find that about $60\%$ of the clusters belonging to the RASS1 Bright Sample
can be confidently defined as extended sources, while
the remaining $\sim 40\%$ are consistent with being pointlike.
If the RASS1 Candidate Sample had been selected on
the basis of source X-ray extent this would have introduced a severe
incompleteness in any flux limited sample.
Indeed even at a hard band count rate limit of 0.25 cts/s (roughly
corresponding to a flux limit of $\sim 3.5\times 10^{-12}$ erg cm$^{-2}$
s$^{-1}$),
almost half of the sources would have been lost.
This is due to the limited angular resolution of the RASS,
in which a significant fraction of the more distant clusters do not appear
as extended sources, coupled to the low photon statistics characterizing
RASS sources.

\section{The Number Counts of the RASS1 Bright Sample}

\subsection{Converting Count Rate to Flux}
To convert a count rate into an X-ray flux in the 0.5-2.0 keV band, 
we assume that all clusters
have a thermal spectrum (Raymond \& Smith model in XSPEC version 9.01)
with a temperature of 5 keV, a metal abundance of 0.5 $Z_{\odot}$ and 
a redshift of 0.1, which is the median redshift of the 
RASS1 Bright Sample.
The assumption of a spectral model and its associated parameters
($T$, $Z$ and $z$) is not critical, as the flux in the 0.5-2.0 keV band 
depends only weakly on these parameters.
We find that by varying them within a range of values that are
typical for clusters of galaxies (T$_{gas} \sim 3-10$ keV, Z $\sim 0.3-1.0$ 
Z$_{\odot}$) and for a reasonable redshift range (z $\leq 0.3$), 
the conversion factor for the ROSAT hard band changes by less than 5\%.

With these assumptions we find that the relation between count rate and flux
as a function of $N_H$ is well fitted, for $N_H$ in the range 
$7\times 10^{19} - 4\times 10^{21}$ cm$^{-2}$, by a quadratic relation 
(see also De Grandi 1996a):
$$S = (1.193 ~ + ~ 3.315 N_H ~ + ~ 2.152 N_H^2) \cdot cr, 
\eqno(1)$$
where $S$ is the 0.5-2.0 keV flux in units of $10^{-11}$ erg cm$^{-2}$
s$^{-1}$, $N_H$ is the Galactic absorption for the individual cluster 
as given in Dickey \& Lockman (1990) in units of $10^{22}$ cm$^{-2}$ 
and $cr$ is the SRT hard band count rate.

\subsection{Sky Coverage and LogN-LogS Distribution}
Our sample has been obtained by performing a cut in count rate.
Since different regions of the sky show different amounts
of Galactic absorption, the cut in count rate translates into a range of
flux limits.
We have computed the flux limit as a function of $N_H$ and from that we have
derived the sky coverage as a function of the flux limit (see Figure 7).
As regions of high Galactic absorption have been excluded from the 
RASS1 Bright Sample ($\S$ 4.1), $N_H$
varies within a limited range, $10^{20} < N_H < 10^{21}$ cm$^{-2}$.
Consequently, the flux limit does not vary much over the available
sky area, so that the sky coverage (see Figure 7) is almost constant
for flux limits larger than $\sim 4\times 10^{-12}$ erg cm$^{-2}$ s$^{-1}$ 
and decreases rapidly to zero at the flux limit of
$3.05\times 10^{-12}$ erg cm$^{-2}$ s$^{-1}$. 

The flux limit should take into account exposure time and in the case of 
clusters the angular extent, e.g., Rosati et al. (1995).
However, in our case both of these dependencies 
may be neglected.
As long as the exposure time is larger than 150 seconds ($\S$ 4.1)
and the extension is less than $\sim$ 4 arcmin ($\S$ 6.1)
a source with an SRT hard band count rate $> 0.25$ cts/s will be detected.

The cumulative LogN-LogS distribution for the 130 objects of the RASS1 
Bright Sample has been computed by summing up the contribution of each 
cluster weighted by the area in which the cluster could have been detected:
$$ N(>S) = \sum_{S_i > S} {1\over \Omega_i},
\eqno(2)$$
where $N(>S)$ is the surface number density of sources with flux larger than 
$S$, $S_i$ is the flux of the $i^{th}$ source and $\Omega_i$ the
associated solid angle. 
The LogN-LogS distribution is plotted in Figure 8.

We have modeled the LogN-LogS distribution with a power law of the form:
$$N(>S) = A S^{-\alpha}, \eqno (3)$$
and computed the power law slope, $\alpha$, using the maximum likelihood
method described in Crawford et al. (1970) and Murdoch et al. (1973), which
uses the unbinned data. 
The likelihood function, $\cal L$, is given in the appendix of Murdoch 
et al. (1973).
The derived value for $\alpha$ ($1.37\pm 0.15$) has
been corrected by the factor $(M-1)/M$ (Crawford et al. 1970), where $M$ 
is the number of objects in the sample, 
and for the bias in the derived slope induced by the presence of measurement 
errors in the fluxes (Murdoch et al. 1973, Table 5).
This last correction is computed by Murdoch et al. (1973) for the case of 
noise-limited flux measurements, however, because of the way our sample 
has been selected, the minimum S/N is not well defined. 
On the basis of an analysis of the overall S/N distribution, we have 
defined an effective limiting S/N ($\sim 8$) corresponding to the peak of 
the distribution and we have applied the correction appropriate for this 
value.
The corrected value of the slope is $\alpha = 1.34\pm 0.15$, where the 
quoted errors are 1-$\sigma$.
The cumulative effect of the two corrections shifts the value
of the slope by an amount which is much smaller than the errors.
We note that the observed slope is consistent within 1-$\sigma$ with the
Euclidean slope. 
The normalization, $A$, has been computed by imposing that the integral 
distribution, $N(>S)$, described by the power law, be equal to the observed 
one at the flux of the weakest cluster in the sample.
The derived value is $A = 11.87\pm 1.04$ sr$^{-1} (10^{-11}$ erg
cm$^{-2}$ s$^{-1})^{\alpha}$.
Since the maximum likelihood method does not establish whether a
model is acceptable, we have applied a Kolmogorov-Smirnov (KS) test to
our data. 
In the case of a power law with slope $\alpha$ equal to the one derived above
the KS test yields a probability of 0.77 for the observed distribution
to be extracted from the parent population, indicating that a power 
law describes adequately our data.

If the best-fit power law is drawn over Figure 8 it appears to be above 
the data for all the fluxes higher than about 
$0.6-0.7\times 10^{-11}$ erg cm$^{-2}$ s$^{-1}$.
This effect is due to the different statistical weight given to the 
data points by the maximum likelihood algorithm used for the fitting 
procedure and to the fact that the distribution reported in Figure 8 
is cumulative.
Indeed by plotting the differential LogN-LogS this effect disappears 
indicating its purely statistical nature.

We have performed an internal consistency test by computing the LogN-LogS
distribution and the best-fit power law for the 119 sources classified 
as certain clusters (A) only.
The best fit parameters, $\alpha = 1.32\pm 0.15$ and A = $11.14\pm 1.01$, 
are in agreement with those found by considering the whole sample.
Moreover, a KS test between the flux distribution of the
sources classified as A and the sources in the B and C groups, 
did not show any evidence of a statistically significant difference between
the two subsamples.

Figure 8 also shows a comparison of our LogN-LogS distribution with
other works.
The points shown in Figure 8 are our data, while the long-dashed box and the
cross represent the EMSS LogN-LogS (Henry et al. 1992) as re-calculated by 
Rosati et al. (1995) and the Piccinotti et al. (1982) point, respectively.
The solid line corresponds to the number counts of the Brightest Cluster
Sample (BCS; Ebeling et al. 1998, Table 2), and the dotted box corresponds 
to the extrapolation of the bright end of the ROSAT Deep Cluster Sample (RDCS)
as given by Rosati et al. (1997).

Our best--fit to the LogN-LogS is systematically above 
that found for the BCS sample.
The difference at a flux of $3.05\times 10^{-12}$ erg cm${^2}$ s$^{-1}$ is 
$19\%$ and is significant at about the 2-$\sigma$ confidence level.
Possible explanations of this difference could be the different efficiencies of
the pre-selection methods used in compiling the two cluster samples
or the different techniques used to estimate the X-ray fluxes.

Another sample of RASS1 clusters already available in the literature, and
covering the sky at $|b_{II}|>20^o$, is the X-ray Brightest Abell Cluster 
sample (XBACs; Ebeling et al. 1996). 
This sample, however, is optically--selected because it was
compiled by looking for cluster soft X-ray emission over the objects in the
Abell and ACO catalogs. Moreover the ROSAT energy band used to
derive the X-ray fluxes was the broad band (0.1-2.4 keV).

In general our data are in good agreement with previous estimates of the 
LogN-LogS distribution, and moreover as both the EMSS and RDCS 
cluster samples are purely X-ray selected, the good agreement in the 
number counts suggests that our selection function, which is partially 
driven by the optical properties of clusters (see $\S$ 3), does not lead 
to a significant incompleteness. 

\section{Completeness of the RASS1 Bright Sample}
In this section we discuss the possible sources of incompleteness in the
RASS1 Bright Sample, which requires a review of the initial selection 
process of the RASS1 Candidate Sample.

\subsection{Biases Introduced by the SASS1 Detection Algorithm
and the SASS1 Count Rate Cut}
As stated in $\S$ 3 the RASS1 Candidate Sample was selected
from the SASS1 source list, and 
therefore one possible origin of incompleteness is related to limitations
of the SASS1 in detecting and characterizing sources.
The detection algorithm implemented within SASS1 was optimized to detect
pointlike sources. 
Both the two sliding window and the ML techniques (see $\S$ 2.1)
use fixed apertures to detect sources, so that extended sources
with low surface brightness may either not be detected or have their count 
rates significantly underestimated.
Another cause of underestimation of source count rates comes from the
assumption made within the ML technique that the source brightness profile is
a sum of Gaussians, which is not a good description of the profile of
most galaxy clusters.

The conservative solution we adopted to effectively limit the 
incompleteness of the RASS1 Bright Sample with respect to extended sources, 
was to choose a very high limiting count rate.
In the following, using the distribution of sources in Figure 4, we try to
estimate the typical angular dimensions of the objects which may have been 
lost due to the effects described above.
From Figure 4 we note that the more extended the source the further 
away it is from the bisector of 
the count rate plane (dot-dashed line), this is because the more extended 
sources suffer a severe underestimation of the count rate by SASS1.
Therefore the sources which were missed in the RASS1 Bright Sample
because they were not detected by SASS1, i.e. those which should fall 
in the top left quadrant in Figure 4, must be highly extended sources.
We have seen that this incompleteness should be limited to $\lesssim 5\%$
($\S$ 4.2.2).
In order to estimate the typical extent of these clusters we have performed 
the following quantitative analysis on the data distribution reported in 
Figure 4.
To first approximation, we can relate the distance of any straight line 
parallel to the one-to-one line with a value of the extension,
i.e. the core radius.
The sources which should fall in the top left quadrant in Figure 4, must 
have an extent larger than that corresponding to the straight line parallel 
to the bisector of the count rate plane crossing the point 
($cr_{SASS1}, cr_{SRT}$) = (0.055, 0.25). 
Drawing this line in Figure 4 we see that 11 sources fall above the line 
and that all these sources have a core radius larger than about 4 arcmin.

Therefore, the $\sim 8.6$ sources that we estimated in $\S$ 4.2.2 to be 
probably missing from the sample because of the flux cuts, very probably 
have a core radius of 4 arcmin or larger.
Taking a value for the physical core radius of 250 kpc and a Hubble constant 
of 50 km s$^{-1}$ Mpc$^{-1}$, we calculate that a cluster with an angular 
core radius of $4$ arcmin would have a redshift of $0.08$.
Thus the $\sim 5\%$ incompleteness discussed in $\S$ 4.2.2 is most likely due
to sources with redshifts smaller than $\sim 0.08$.
This should be treated as a first order estimate, first because we have used
an approximate analysis and second because clusters show a scatter in their 
physical core radii.

From a preliminary analysis of the ACO clusters within the area of the 
RASS1 Bright Sample (B\"ohringer et al. in preparation), 
we find that 6 ACO clusters with hard band count rate larger than 0.25 cts/s 
are missing from our sample.
All but one of these clusters were detected from SASS1, but their SASS1 
broad band count rates were $< 0.055$ cts/s, i.e. below the initial SASS1 
count rate limit of the RASS1 Candidate Sample.
All these clusters have a measured redshift $\lesssim 0.06$, in agreement with
our expectations about the bias against the more extended and
nearby clusters and our completeness estimate of $\S 4.2.2$. 
The properties of these 6 ACO clusters are reported in Table 3.

{\it Column (1). ---} Name of the ACO cluster.

{\it Column (2). ---} Right ascension (J2000) in degrees. 

{\it Column (3). ---} Declination (J2000) in degrees.

{\it Column (4). ---} SASS1 count rate measured in the (0.1-2.4 keV) band.

{\it Column (5). ---} measured redshift.

{\it Column (6). ---} redshift reference.

\subsection{Biases Introduced by the Identification Process} 

As described in $\S$ 3, the RASS1 Candidate Sample was selected
by basically two means of identification:
the clusters were either found directly by their optical
counterparts (overdensities in the COSMOS galaxy
catalog -- either direct or pre-processed --, or optical clusters 
in the Abell and ACO catalogs), or by having been flagged 
as extended sources during the SASS1 source analysis.  

Clearly, this procedure is potentially prone to a number of selection
effects, that can be summarized as follows.

1) {\it Optical counterparts}.  The COSMOS galaxy catalog was produced 
through the analysis of digitized ESO/SRC J survey optical plates 
(Yentis et al. 1992), using automatic algorithms.  
One recognized problem of the digitization
and star--galaxy separation processes
is in the correct treatment of diffraction spikes and halos 
around bright stars and of the extended envelopes of cD galaxies
(Heydon-Dumbleton et al. 1989).  For our purposes,
one can reasonably think that this effect will in general increase the 
contamination of the RASS1 Candidate Sample, 
by including spurious ``clusters'',
but should not reduce its completeness.   A more serious concern
is the misclassification of galaxies as stars, estimated to be
around $5\%$ in the COSMOS data.  This could potentially reduce the contrast 
of a poor cluster and thus exclude it from the sample.

Concerning, on the other hand, the use of the Abell and ACO catalogs,
one might worry that they are biased against poor systems (expected to
be anyway rare in our sample, at $z$ larger than $\sim 0.04$), and are 
affected by subjective biases that are difficult to quantify a priori.  

2) {\it X--ray extension}.  Sources were classified as extended using 
the threshold values, 
extent radius larger than
25 arcsec and extent likelihood larger than 7 as supplied
by the ML algorithm of SASS1.  This set of cuts has been used
successfully before (for details see e.g. Fig. 7 in Ebeling et al. 1993).
This method is however not reliable in recovering 
all extended sources, as we find directly that several very extended 
sources are not recognized.

To better understand the completeness of the global identification process
within the selection limits of the RASS1 Bright Sample, we have tried to 
exploit the complementarity of these two methods.  (Note that most of the 
clusters in the sample have been found by both methods).

Let us assume that the two means of identification
are uncorrelated, which is the case if clusters are missed
by simple independent errors in the two techniques.  
This allows us to statistically infer the
incompleteness of the sample in the following way. We define 
by $O$ the set of clusters found by optical means and by $X$
the set found by X-ray extent.

The statistical independence of the two search methods, allows us to 
write the combined probability of events $O$ and $X$, as 
$$P(O\cap X) = P(O)\cdot P(X).
\eqno(4)$$
Defining $T$ as the parent sample (i.e., $P(T)\equiv 1$), we can relate
probabilities, $P$, to occurrences, $N$, in the following way:
$P(X)=N(X)/N(T)$, $P(O)=N(O)/N(T)$ and $P(O \cap X)=N(O \cap X)/N(T)$.
Substituting in equation (4) we obtain:
$$N(T) = {{N(O)~ N(X)}\over {N(O \cap X)}},
\eqno(5)$$
and using the actual numbers in the subsets, $N(O) = 118$
$N(X) = 95$, and $N(O \cap X) = 83$, we find $N(T) = 135$.
Consequently, the missing fraction of objects for our 130 clusters 
sample is $\sim 3.7\%$.  
 
The assumption that the two methods of detection are uncorrelated is 
probably not strictly valid, as poor and more distant
clusters will be harder to find for  both techniques. Therefore
the number calculated for the clusters missed is 
a lower limit to the missing fraction, but it 
is already reassuring that this fraction is as low as 3.7\%.

\subsection{$<V/V_{max}>$ Test}
As a final check, we have tested the spatial distribution of the bright
clusters with the $V/V_{max}$ method.
Since our sample has different flux limits ($\S$ 5.2) we have used the
generalization of the $V/V_{max}$ method (Schmidt 1968) given in
Avni \& Bahcall (1980).
We have also assumed an Einstein-deSitter cosmological model with
$\Lambda = 0$, $q_0 = 0.5$ and $H_0 = 50$ km s$^{-1}$ Mpc$^{-1}$.
The derived $<V/V_{max}>$ is $0.49\pm 0.16$, consistent with uniformity.

\section{Summary and Conclusions}

The aim of the present work was to derive an X-ray flux limited sample
of bright clusters of galaxies characterized by a high degree of
completeness.
It is based on a first cluster candidate sample derived in the ESOKP
collaboration, belonging to the southern Galactic cap region ($\delta < 2.5^o$
and $b_{II} < -20^o$).
This is called in our paper the RASS1 Candidate Sample and
contains 679 sources.

We performed first a detailed reanalysis of fluxes for all sources  
in this RASS1 Candidate Sample by using the RASS merged data and 
the SRT method developed and discussed in Paper I. 

We have applied to the RASS1 Candidate Sample various restrictive
selections aimed at heavily reducing the sources of incompleteness.
Our first selection set a lower limit of 150 s to the exposure time in order 
to avoid regions of sky where objects could have been missed because of 
the low sensitivity of the survey.
In the second selection we excluded crowded regions of the sky, i.e. 
the Galactic plane and the Magellanic Clouds, to avoid confusion problems 
affecting optical and X-ray catalogs.
The third selection set a lower limit to the SRT count rate of 0.25 cts/s
in the hard band.
In setting the SRT count rate limit, we investigated the behavior of a 
control sample of optically identified objects, namely the EMSS sample 
reobserved in the RASS.
These selections yielded an X-ray completeness in the derived sample of 
$\sim 95\%$. 
Such a high degree of completeness was reached at the expenses of reducing 
drastically the number of candidates in our sample, from 679 to 164.

We used our new data from the ESOKP, together with data drawn from
the literature to identify the selected candidates.
After removing a number of false identifications we produced
a sample which contains 130 clusters with X-ray fluxes larger than 
$\sim 3.5\times 10^{-12}$ erg cm$^{-2}$ s$^{-1}$ and $z\lesssim 0.3$, 
covering a sky area of 8235 deg$^2$.

We have then investigated the various sources of incompleteness 
and biases, which could be affecting the RASS1 Bright sample.
The key factor allowing us to constrain the incompleteness has been to
apply a cut at a relatively high X-ray count rate.
This action limits the bias against very extended X-ray sources (i.e. 
nearby clusters and groups).
From our estimates we have also seen that this bias is kept under control 
for redshifts $\lesssim 0.08$ and eliminated for redshifts $\gtrsim 0.08$.

A statistical estimate of the completeness level of the identification
procedure, based on the relative success rates of the two main methods 
of cluster identification (optical- vs. X-ray-based), indicates an
incompleteness $\sim 4\%$ due to this source.  Adding this to that
estimated from the flux selection procedure, we obtain a global 
completeness for the sample which is better than $90\%$.

The LogN-LogS distribution is well described 
by a power law with slope $\alpha = 1.34\pm 0.15$ and normalization 
$A = 11.87\pm 1.04$ sr$^{-1} (10^{-11}$ erg cm$^{-2}$ s$^{-1})^{\alpha}$.
A comparison between our result and previous measurements shows good 
agreement.

The sample discussed here should represent a useful database 
for a number of statistical studies on the properties of 
clusters of galaxies in the local Universe.
Finally, we should mention that the results presented here will be
extended by the future developments of the ongoing ESOKP collaboration.

\acknowledgments

The authors would like to thank the ROSAT team at MPE. 
H. T. MacGillivray at ROE and D. J. Yentis at NRL for having provided 
the COSMOS digitized optical sky survey catalogs.
We thank also K. Romer for having allowed us to publish eight redshifts 
from the SGP project and J. P. Henry and C. R. Mullis for having provided 
two new redshifts.  
SDG would like to thank P. Rosati, R. Della Ceca and G. Zamorani for
useful comments and discussions.
Special thanks also to C. Izzo at MPE for his help in the data analysis.
This research has made use of data provided by the NED and the SIMBAD 
databases and by the digitized optical images from the POSS and UK Schmidt 
sky surveys obtained through the SKYVIEW facility.
We also used data from the HEASARC online service and the ROSAT public 
archive.
We thank all those who contribute in maintaining these databases.


\clearpage

\begin{figure}
\plotone{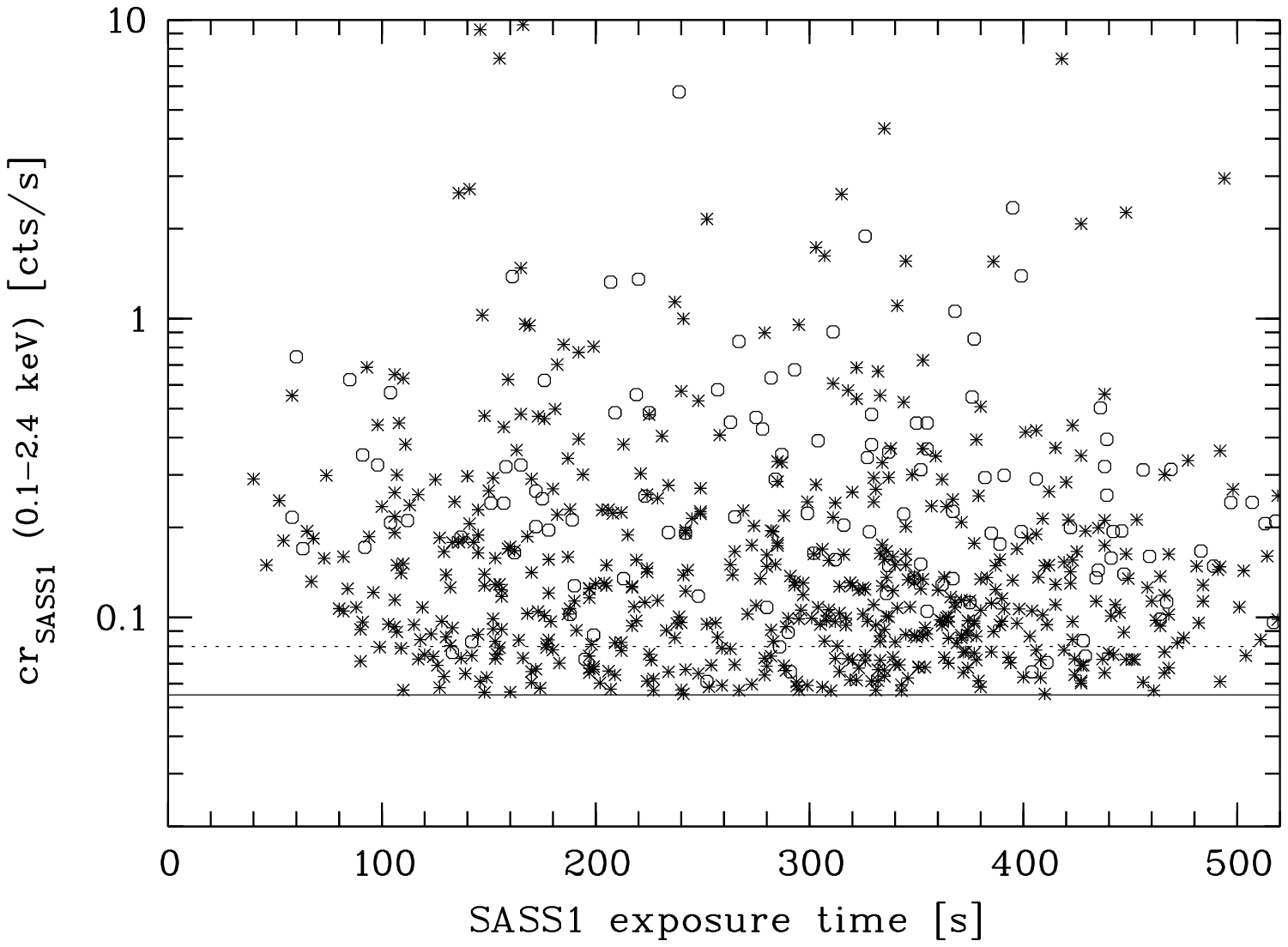}
\vskip -11cm
\caption 
{SASS1 broad band count rates versus SASS1 exposure times for the
RASS1 Candidate Sample. Sources which were found to be pointlike with
the SRT are indicated as asterisks, while sources which were found to be 
extended are indicated as open circles. The dotted and solid lines represent 
the limiting SASS1 count rates 0.08 and 0.055 cts/s, respectively.}
\end{figure}
\clearpage

\begin{figure}
\plotone{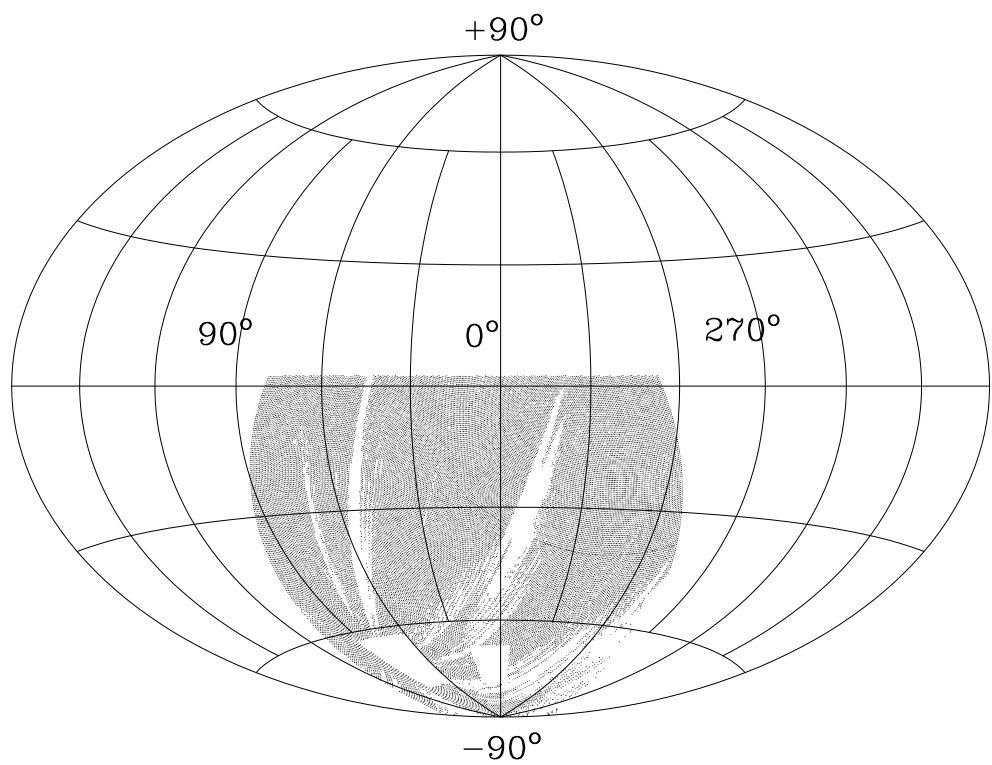}
\vskip -11cm
\caption 
{Total sky area covered by the RASS1 Bright Sample in equatorial 
coordinates (J2000.0).
Regions with low exposure times ($< 150$ s) and the Magellanic 
Clouds are shown in white.}
\end{figure}
\clearpage

\begin{figure}
\plotone{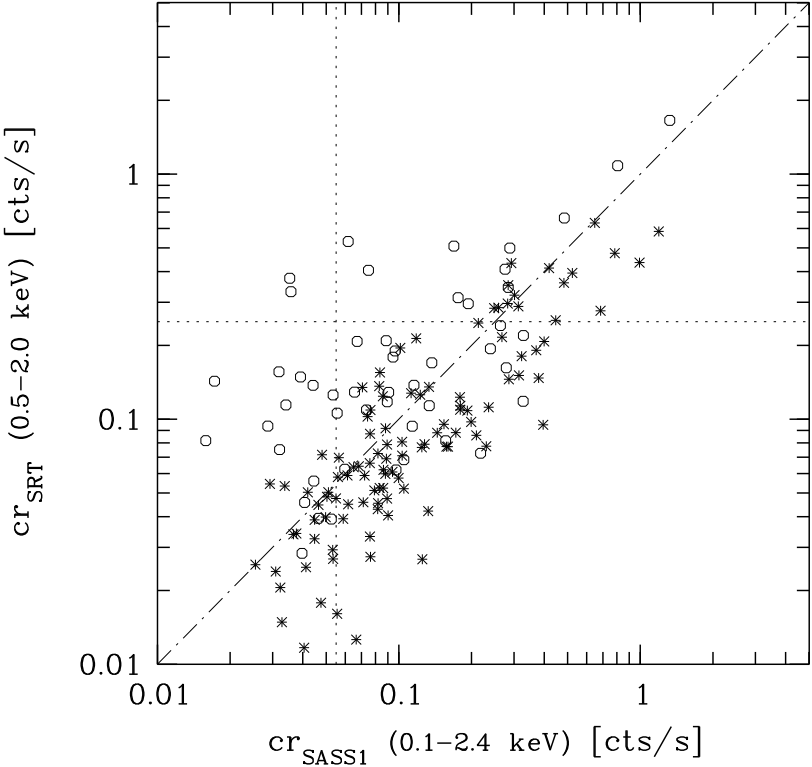}
\vskip -11cm
\caption 
{SRT hard band count rates versus SASS1 broad band count rates for
the EMSS control sample. Asterisks indicate pointlike objects (i.e. AGNs and
stars), open circles indicate potentially extended sources 
(i.e. galaxies and galaxy clusters).
The vertical and horizontal dotted lines represent the SASS1 count rate limit, 
0.055 cts/s, and the SRT count rate limit, 0.25 cts/s, respectively.
The dot-dashed line indicates the bisector of the count rate plane.}
\end{figure}
\clearpage

\begin{figure}
\plotone{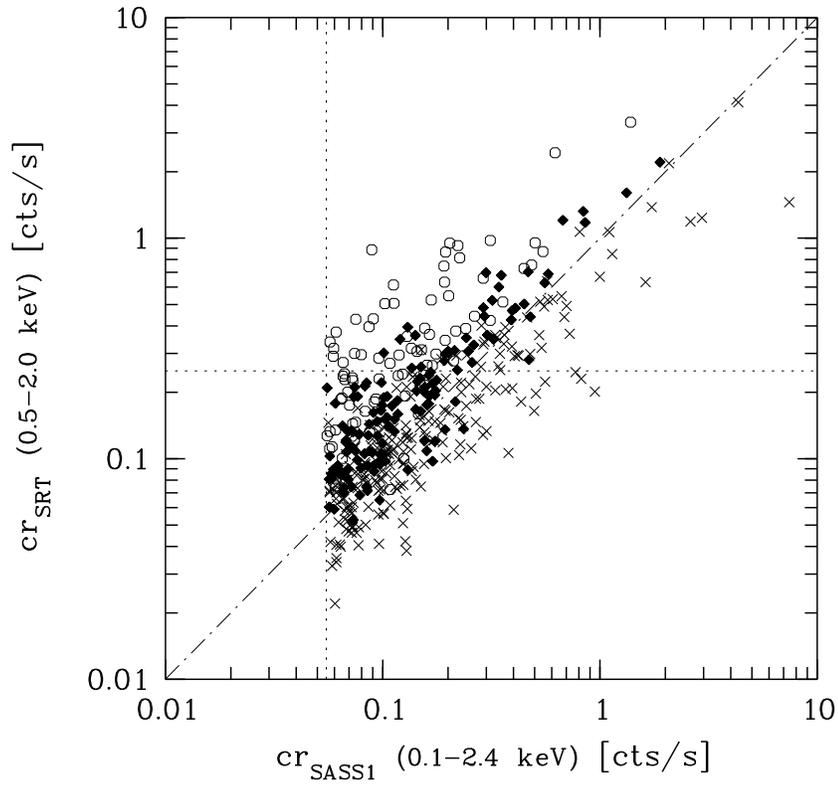}
\vskip -11cm
\caption 
{SRT hard band count rates versus SASS1 broad band count rates for
the SUB1 cluster candidate sample (see $\S$ 4.2.2 for definition). 
Crosses correspond to sources with core radii $<$ 1 arcmin, filled squares
to sources with core radii from 1 to 2 arcmin, and open circles to 
sources with core radii $>$ 2 arcmin.
The vertical and horizontal dotted lines represent the SASS1
count rate limit, 0.055 cts/s, and the SRT count rate limit,
0.25 cts/s, respectively.}
\end{figure}
\clearpage

\begin{figure}
\plotone{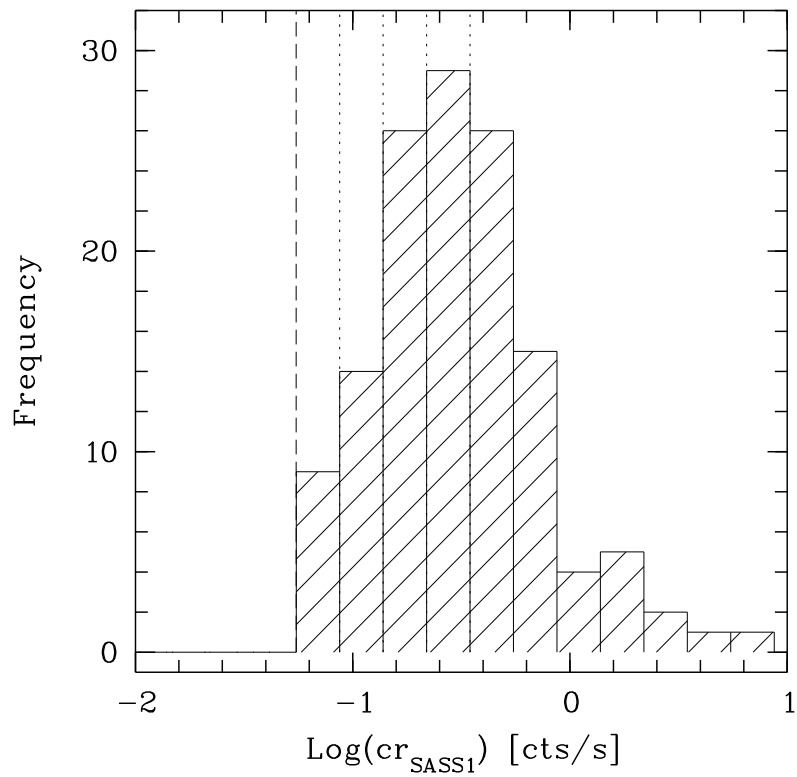}
\vskip -11cm
\caption 
{SASS1 broad band count rates of SUB1 sources with an SRT hard band 
count rates $> 0.25$ cts/s. The long dashed line is the SASS1 count
rate limit of 0.055 cts/s, the dotted lines show the bins used for the
fit (see text for details).}
\end{figure}
\clearpage

\begin{figure}
\plotone{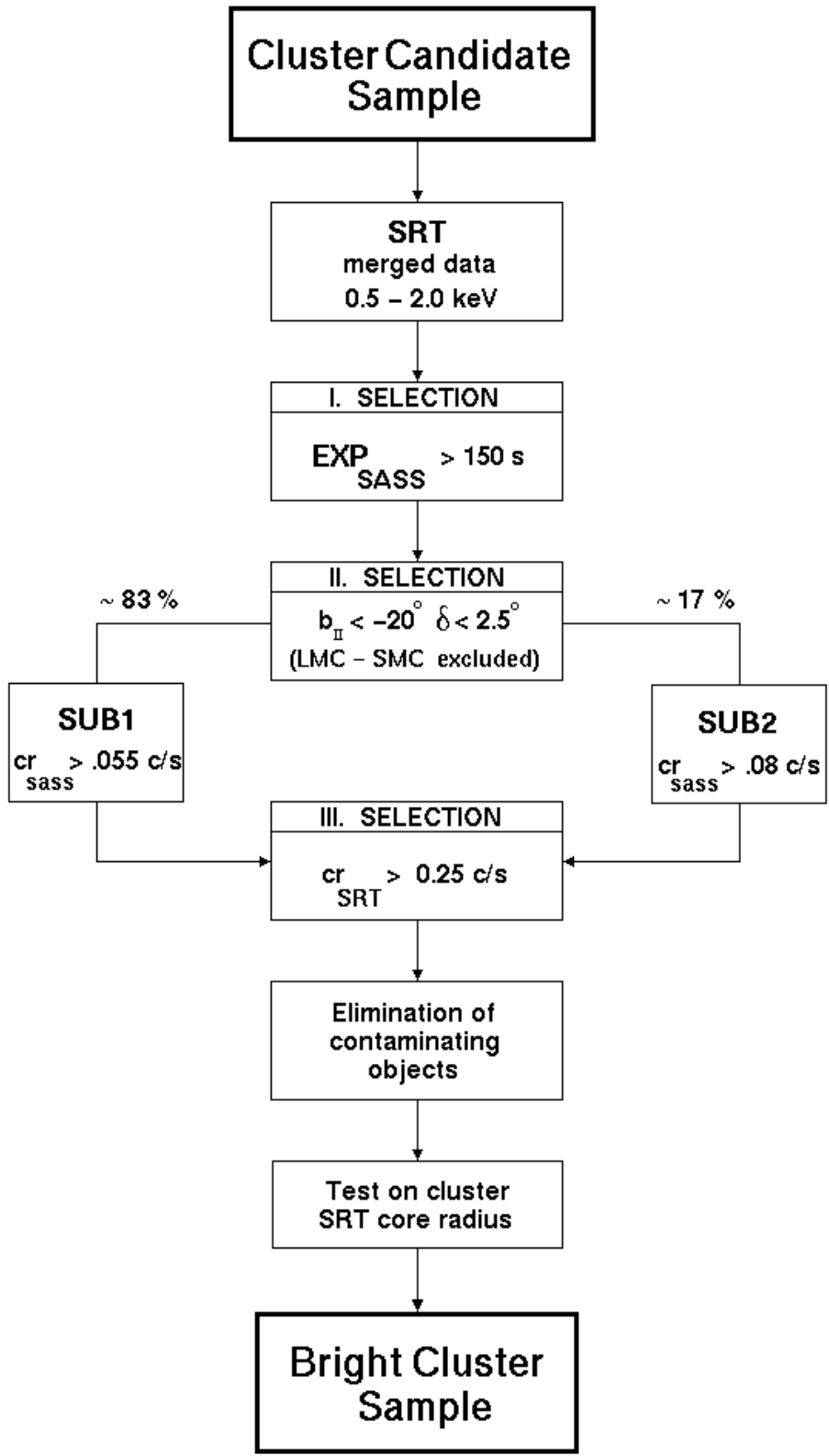}
\caption 
{Schematic representation of the selections applied to the
RASS1 Candidate Sample to obtain the RASS1 Bright Sample.}
\end{figure}
\clearpage

\begin{figure}
\plotone{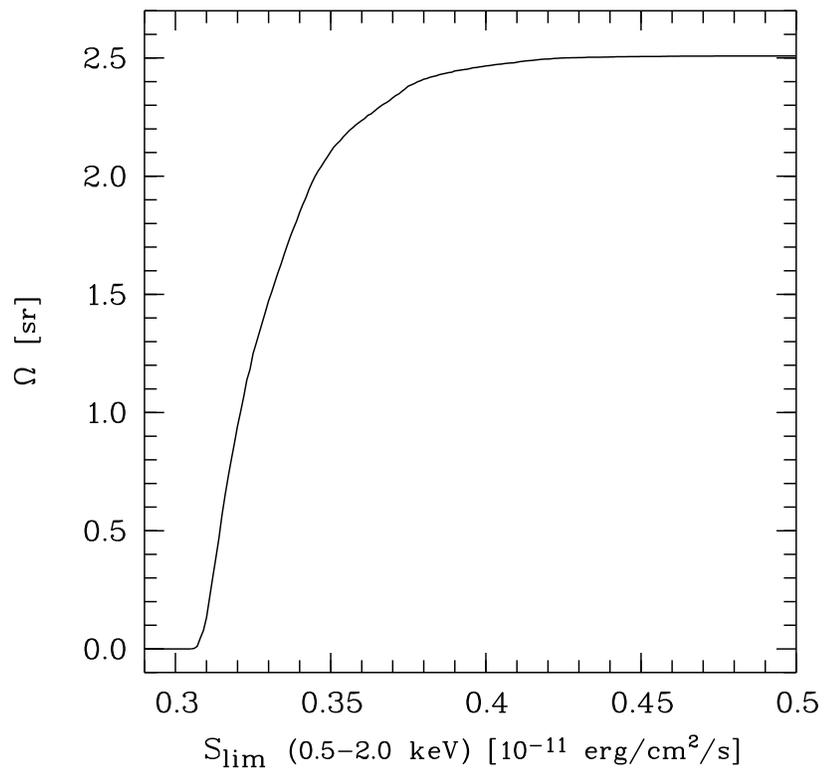}
\vskip -11cm
\caption 
{Sky coverage as a function of the flux limit for the 
RASS1 Bright Sample.}
\end{figure}
\clearpage

\begin{figure}
\plotone{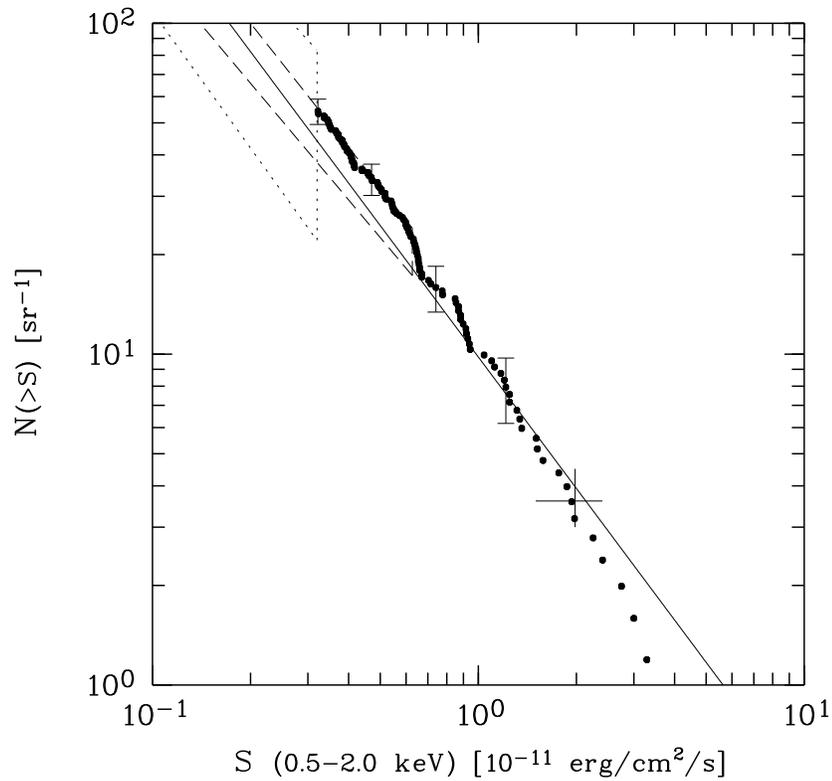}
\vskip -11cm
\caption 
{Cumulative cluster number counts distribution, LogN-LogS,
of the RASS1 Bright Sample (dots).
Vertical error bars on a few individual points represent the uncertainty in 
the number of clusters.
Also shown for comparison are the LogN-LogS of other cluster samples:
the long-dashed box represents the EMSS LogN-LogS (Henry et al. 1992) as 
recalculated in Rosati et al. (1995), the cross represents the Piccinotti 
et al. (1982) point, the dotted box is the extrapolation of the bright end 
of the RDCS sample by Rosati et al. (1997), and the solid line represents
the best-fit of the BCS sample (Ebeling et al. 1998).} 
\end{figure}
\clearpage

\textheight=10.in
\begin{deluxetable}{crcrcccccccr}
\tablewidth{0pc}
\tablefontsize{\footnotesize}
\tableheadfrac{0.01}
\tablecaption{Cluster Catalog}
\tablehead{
\colhead{Sequence}&\colhead{RA(J2000)}&\colhead{NH}&\colhead{Time}&\colhead{cr(5$^{\prime}$)}&\colhead{cr(SRT)}&
\colhead{P}&\colhead{F$_{X}$}&\colhead{L$_{X}$}&\colhead{ID}&\colhead{z}&\colhead{Note}\nl
\colhead{ }&\colhead{Dec(J2000)}&  &  &\colhead{Err} &\colhead{Err $\pm$}&  &\colhead{Err}&\colhead{Err}&
\colhead{Class}& \colhead{Ref}&\colhead{ }  \nl
\colhead{(1)}&\colhead{(2)}&\colhead{(3)}&\colhead{(4)}&\colhead{(5)}&\colhead{(6)}&\colhead{(7)}&\colhead{(8)}&\colhead{(9)}& 
\colhead{(10)}&\colhead{(11)}&\colhead{(12)}}
\startdata
001 & 00 03 09.41 & 1.09 & 335.7 & 0.26 & 0.51 & 9.9e$-$04 & 0.62 & 0.64           &A2717 & 0.0490    &  \nl
 &$-$35 56 22.0 &      &        & 0.03 & $+$0.13 $-$0.10 &             & 0.14 & 0.15     & A  & (1) & \nl \nl
002 & 00 05 59.39 & 1.16 & 348.4 & 0.20 & 0.27 & 4.0e$-$02 & 0.34 & 1.86           &A2721 & 0.1140    &  \nl 
 &$-$34 43 05.5 &      &        & 0.03 & $+$0.05 $-$0.05 &             & 0.06 &    0.33   &A  & (2)  &  \nl \nl
003 & 00 11 20.51  & 1.84 &  329.5 & 0.40 & 0.70  &  3.6e$-$04 & 0.88 &    1.44  &A2734    & 0.0617 &   \nl 
 &$-$28 51 05.0 &      &        & 0.04 & $+$0.12 $-$0.11 &             & 0.14 &    0.23   &A  & (1)   &  \nl \nl
004 & 00 13 37.12  & 2.00 &  318.1 & 0.22 & 0.30  &  4.9e$-$02 & 0.37 &   1.41   &A0013    & 0.0943 &   \nl 
 &$-$19 29 54.0 &      &        & 0.03 & $+$0.05 $-$0.05 &             & 0.06 &    0.24   &A  & (1)  &  \nl \nl
005 & 00 14 20.32  & 1.65 &  332.2 & 0.23 & 0.28  &  1.1e$-$01 & 0.35 &   13.97  &A2744    & 0.3080 &   \nl 
 &$-$30 22 58.0 &      &        & 0.03 & $+$0.04 $-$0.04 &             & 0.05 &    2.16   &A  & (2)  &  \nl \nl
006 & 00 20 44.05  & 2.26 &  318.7 & 0.21 & 0.33  &   1.5e$-$02 & 0.42 &    3.62  &A0022   & 0.1432 &   \nl 
 &$-$25 42 21.0 &      &        & 0.03 & $+$0.07 $-$0.06 &             & 0.08 &    0.72   &A  & (3)  &  \nl \nl
007 & 00 25 34.36  & 1.69 &  337.8 & 0.29 & 0.51  &  1.1e$-$03 & 0.64 &    0.68  &S0041    & 0.0498 &   \nl
 &$-$33 02 43.0 &      &        & 0.03 & $+$0.11 $-$0.09 &             & 0.12 &    0.13   &A  & (2)  &  \nl \nl
008 & 00 28 35.89  & 1.82 &  313.2 & 0.21 & 0.39  &  4.1e$-$03 & 0.49 &    2.46  &A0042    & 0.1087 &   \nl 
 &$-$23 39 07.0 &      &        & 0.03 & $+$0.12 $-$0.09 &             & 0.13 &    0.65   &A  & (4)  &  \nl \nl
009 & 00 41 50.11  & 3.58 &  385.8 & 1.89 & 3.11  &  3.6e$-$07 & 4.09 &    5.41  &A0085    & 0.0556 &   \nl
 &$-$09 18 17.5 &      &        & 0.07 & $+$0.19 $-$0.19 &             & 0.25 &    0.33   &A  & (1)  &  \nl \nl
010 & 00 42 08.63  & 1.49 &  336.0 & 0.41 &  0.49  &  1.2e$-$01 & 0.61 &    3.06  &A2811   & 0.1087 &   \nl 
 &$-$28 32 09.0 &      &        & 0.04 & $+$0.05 $-$0.05 &             & 0.07 &    0.33   &A  & (5)  &  \nl  \nl
011 & 00 49 24.05  & 1.80 &  331.5 & 0.19 &  0.26  &  5.2e$-$02 & 0.32 &    1.66  &S0084   & 0.1100 &   \nl 
 &$-$29 31 21.0 &      &        & 0.03 & $+$0.05 $-$0.05 &             & 0.06 &    0.32   &A  & (2)  &  \nl \nl
012 & 00 52 34.76  & 6.64 &  252.2 & 0.30 &  0.36  &  1.5e$-$01 & 0.52 &    2.87  &A2837   & 0.1142 &   \nl 
 &$-$80 15 21.0 &      &        & 0.04 & $+$0.06 $-$0.05 &             & 0.08 &    0.43   &A  & (6)  &  \nl \nl
013 & 00 56 11.69  & 3.10 &  295.6 & 0.43 &  1.85  &  1.8e$-$06 & 2.41 &   2.90  &A0119    & 0.0442 &   \nl 
 &$-$01 14 52.5 &      &        & 0.04 & $+$0.66 $-$0.55 &             & 5.16 &    4.32   &A  & (1)  &  \nl \nl
014 & 01 02 42.21  & 1.60 &  361.5 & 0.73 &  1.27  &  1.7e$-$05 & 1.58 &    2.16  &A0133   & 0.0566 &   \nl 
 &$-$21 52 43.5 &      &        & 0.05 & $+$0.15 $-$0.13 &             & 0.17 &    0.24   &A  & (1)  &  \nl \nl
015 & 01 07 49.61  & 1.86 &  388.3 & 0.16 & 0.30  &  6.4e$-$03 & 0.37 &    2.35  &A2871    & 0.1219 &   \nl
 &$-$36 43 43.5 &      &        & 0.02 & $+$0.09 $-$0.07 &             & 0.10 &    0.64   &A  & (1)  &  \nl \nl
016 & 01 08 13.14  & 3.01 &  434.8 & 0.14 &  0.40  &  1.5e$-$03 & 0.52 &    0.42  &A0147   & 0.0438 &   \nl 
 &$-$02 11 33.0 &      &        & 0.02 & $+$0.25 $-$0.13 &             & 0.26 &    0.21   &A  & (4)  &  \nl \nl
017 & 01 08 52.89  & 1.69 &  460.2 & 0.25 &  0.51  &  1.9e$-$04 & 0.63 &    0.77  &A0151   & 0.0533 &   \nl
 &$-$15 25 45.5 &      &        & 0.02 & $+$0.11 $-$0.09 &             & 0.13 &    0.16   &A  & (1)  &  \nl \nl
018 & 01 09 52.59  & 2.10 &  310.9 & 0.21 &  0.39  &  4.4e$-$03 & 0.49 &    0.12  &A2877   & 0.0241 &   \nl 
 &$-$45 55 42.0 &      &        & 0.03 & $+$0.11 $-$0.09 &             & 0.13 &    0.03   &A  & (2)  &  \nl \nl
019 & 01 14 59.52  & 3.32 &  434.5 & 0.17 & 0.71  &  2.6e$-$04 & 0.93 &    0.97  &A0168    & 0.0448 &   \nl 
 &$-$00 22 13.9 &      &        & 0.02 & $+$0.26 $-$0.22 &             & 2.23 &    1.92   &A  & (7)  &  \nl \nl
020 & 01 20 58.78  & 1.85 &  293.7 & 0.39 & 0.73  &  3.1e$-$04 & 0.92 &   1.03  &0118.5-1408 & 0.0511 & 1 \nl 
 &$-$13 51 07.0 &      &        & 0.04 & $+$0.14 $-$0.12 &             & 0.17 &    0.19  &A  &  (8)   &  \nl \nl
021 & 01 25 29.89  & 3.08 &  435.1 & 0.19 & 0.30  &  7.0e$-$03 & 0.39 &    0.05  &NGC0533  & 0.0171 & 2 \nl 
 &$-$01 45 40.0 &      &        & 0.02 & $+$0.06 $-$0.05 &             & 0.07 &    .01   &A  & (6) &  \nl \nl
022 & 01 31 52.69  & 1.56 &  466.4 & 0.24 & 0.33  &  1.2e$-$02 & 0.42 &    7.43  &A0209    & 0.2060 &   \nl 
 &$-$13 36 38.0 &      &        & 0.02 & $+$0.05 $-$0.05 &             & 0.06 &    1.06  &A  & (4)  &  \nl \nl
023 & 01 37 16.19  & 2.75 &  456.7 & 0.26 & 0.51  &  1.7e$-$04 & 0.66 &    0.44  & \nodata  & 0.0392 & 3 \nl
 &$-$09 11 40.0 &      &        & 0.03 & $+$0.11 $-$0.09 &             & 0.13 &    0.08  &A  & (8)   &  \nl \nl
024 & 01 45 07.63  & 2.35 &  251.2 & 0.22 & 0.43  &  5.0e$-$03 & 0.55 &    3.35  &A2941    & 0.1183 &   \nl 
 &$-$53 01 58.5 &      &        & 0.03 & $+$0.09 $-$0.09 &             & 0.17 &    1.03  &A  & (6)  &  \nl \nl
025 & 02 25 52.81  & 2.14 &  409.3 & 0.18 & 0.28  &  7.7e$-$03 & 0.35 &    7.43  &A3017   & 0.2195 & 4 \nl
 &$-$41 54 46.0 &      &        & 0.02 & $+$0.04 $-$0.04 &             & 0.08 &    1.53   &A  & (8)   &  \nl \nl
026 & 02 31 56.22  & 2.89 &  256.0 & 0.08 & 0.29  &  2.3e$-$02 & 0.38 &    0.08  &UGC02005 & 0.0221 & 5 \nl 
 &$-$01 14 59.5 &      &        & 0.02 & $+$2.24 $-$0.17 &             & 2.05 &    0.43   &C  & (8)   &  \nl \nl
027 & 02 32 18.69  & 2.61 &  388.8 & 0.22 & 0.31  &  2.0e$-$02 & 0.40 &   13.29  & \nodata  & 0.2836 & 6 \nl 
 &$-$44 20 41.5 &      &        & 0.03 & $+$0.05 $-$0.05 &             & 0.07 &    2.17   &A  & (8)   &  \nl \nl
028 & 02 49 36.03  & 1.80 &  430.7 & 0.23 & 0.37  &  4.1e$-$03 & 0.46 &    0.10  &S0301    & 0.0223 &   \nl 
 &$-$31 11 09.5 &      &        & 0.02 & $+$0.06 $-$0.06 &             & 0.08 &    0.02   &A  & (2)  &  \nl \nl
029 & 03 03 24.17  & 7.82 &  259.7 & 0.18 & 0.26  &  4.5e$-$02 & 0.38 &    3.79  &A0409    & 0.1530 &   \nl 
 &$-$01 56 03.5 &      &        & 0.03 & $+$0.06 $-$0.06 &             & 0.09 &    0.85   &A  & (9)  &  \nl \nl
030 & 03 07 03.50  & 1.36 &  254.3 & 0.18 & 0.26  &  2.9e$-$02 & 0.32 &    9.15  &A3088    & 0.2534 &   \nl 
 &$-$28 40 01.5 &      &        & 0.03 & $+$0.05 $-$0.05 &             & 0.09 &    2.31   &A  & (6) &  \nl \nl
031 & 03 14 22.59  & 3.57 &  345.7 & 0.29 & 0.46  &  3.7e$-$03 & 0.60 &    1.32  &A3104    & 0.0718 &   \nl
 &$-$45 25 10.5 &      &        & 0.03 & $+$0.08 $-$0.07 &             & 0.10 &    0.22   &A  & (6) &  \nl \nl
032 & 03 17 58.85  & 2.53 &  513.0 & 1.07 & 1.38  &  1.5e$-$02 & 1.77 &    4.24  &A3112    & 0.0750 &   \nl 
 &$-$44 14 07.0 &      &        & 0.05 & $+$0.08 $-$0.08 &             & 0.10 &    0.24   &A  & (1)  &  \nl \nl
033 & 03 28 37.24  & 3.09 &  559.7 & 0.31 & 0.44  &  5.9e$-$03 & 0.57 &    1.79  &A3126    & 0.0856 &   \nl
 &$-$55 42 27.5 &      &        & 0.02 & $+$0.05 $-$0.05 &             & 0.06 &    0.20   &A  & (1)  &  \nl \nl
034 & 03 30 01.11  & 1.47 &  764.1 & 0.22 & 0.44  &  2.2e$-$05 & 0.55 &    0.72  &A3128    & 0.0554 & 7 \nl 
 &$-$52 35 39.5 &      &        & 0.02 & $+$0.08 $-$0.07 &             & 0.09 &    0.12   &A  & (2)  &  \nl \nl
035 & 03 42 53.06  & 1.06 &  702.9 & 0.89 & 1.83  &  1.6e$-$11 & 2.25 &    3.36  &A3158    & 0.0591 &   \nl 
 &$-$53 37 43.0 &      &        & 0.04 & $+$0.17 $-$0.15 &             & 0.20 &    0.29   &A  & (1)  &  \nl \nl
036 & 03 45 58.30  & 1.58 &  484.2 & 0.21 & 0.35  &  2.0e$-$03 & 0.44 &    2.07  &A0458    & 0.1050 &   \nl 
 &$-$24 16 45.0 &      &        & 0.02 & $+$0.07 $-$0.06 &             & 0.08 &    0.36   &A  & (4)  &  \nl \nl
037 & 03 51 25.86  & 7.65 &  368.0 & 0.31 & 0.85  &  3.2e$-$05 & 1.25 &    2.00  &S0405    & 0.0613 &   \nl 
 &$-$82 12 44.5 &      &        & 0.03 & $+$0.32 $-$0.21 &             & 0.40 &    0.64   &A  & (6)  &  \nl \nl
038 & 03 52 25.09  & 8.02 &  766.9 & 0.38 & 0.71  &  4.0e$-$07 & 1.04 &    7.62  &A3186    & 0.1270 &   \nl 
 &$-$74 01 02.5 &      &        & 0.02 & $+$0.12 $-$0.11 &             & 0.14 &    0.96   &A  & (10)  &  \nl \nl
039 & 04 13 59.19  & 1.41 &  481.8 & 0.42 & 0.63  &  1.5e$-$03 & 0.78 &    0.84  &1ES0412-382 & 0.0502 & 8 \nl
 &$-$38 05 50.0 &      &        & 0.03 & $+$0.07 $-$0.07 &             & 0.08 &    0.09   &A  & (6) &  \nl \nl
040 & 04 19 39.01  & 11.5 &  291.5 & 0.73 & 1.20  &  1.3e$-$04 & 1.93 &    0.13  &NGC1550  & 0.0123 & 9 \nl 
 &$-$02 24 24.5 &      &        & 0.05 & $+$0.14 $-$0.13 &             & 0.22 &    0.01   &B  & (11)  &  \nl \nl
041 & 04 25 51.02  & 6.40 &  272.7 & 0.88 & 1.32  &  7.2e$-$04 & 1.87 &    1.26  &EXO0422-086 & 0.0397 &  \nl
 &$-$08 33 38.5 &      &        & 0.06 & $+$0.13 $-$0.13 &             & 0.18 &    0.12   &A  & (6) &  \nl \nl
042 & 04 30 58.82  & 1.48 & 1460.6 & 0.75 & 2.42  &  1.3e$-$33 & 3.00 &    9.53  &A3266    & 0.0589 &   \nl 
 &$-$61 27 52.5 &      &        & 0.02 & $+$0.70 $-$0.58 &             & 1.92 &    2.85   &A  & (1)  &  \nl \nl
043 & 04 33 37.07  & 5.68 &  251.1 & 1.82 & 3.35  &  6.9e$-$08 & 4.65 &    2.15  &A0496    & 0.0328 &   \nl 
 &$-$13 15 20.0 &      &        & 0.09 & $+$0.31 $-$0.28 &             & 0.41 &    0.19   &A  & (7)  &  \nl \nl
044 & 04 45 11.28  & 4.78 &  492.3 & 0.19 & 0.65  &  9.6e$-$05 & 0.88 &    0.50  &NGC1650  & 0.0363 & 10 \nl 
 &$-$15 51 14.0 &      &        & 0.02 & $+$0.42 $-$0.22 &             & 0.46 &    0.26   &B  & (12)  &  \nl \nl
045 & 05 00 43.81  & 3.07 &  456.0 & 0.18 & 0.50  &  3.2e$-$04 & 0.65 &    0.80  &A3301    & 0.0536 &   \nl 
 &$-$38 40 25.5 &      &        & 0.02 & $+$0.25 $-$0.15 &             & 0.26 &    0.33   &A  & (1)  &  \nl \nl
046 & 05 10 43.85  & 8.30 &  406.3 & 0.22 & 0.31  &  2.4e$-$02 & 0.46 &    9.30  &\nodata & 0.2195 &   \nl 
 &$-$08 01 13.0 &      &        & 0.03 & $+$0.05 $-$0.05 &             & 0.07 &    1.47   &A  & (6) &  \nl \nl
047 & 05 25 32.56  & 1.75 &  517.0 & 0.31 & 0.54  &  2.5e$-$04 & 0.67 &    0.41  &A3341    & 0.0378 &   \nl
 &$-$31 36 13.5 &      &        & 0.03 &  $+$0.08 $-$0.07 &             & 0.10 &    0.06   &A  & (1)  &  \nl \nl
048 & 05 28 55.06  & 2.10 &  638.2 & 0.24 & 0.30  &  1.1e$-$01 & 0.37 &   12.54  & \nodata & 0.2839 & 11 \nl 
 &$-$39 27 54.0 &      &        & 0.02 & $+$0.03 $-$0.03 &             & 0.04 &    1.33   &A  & (6) &  \nl \nl
049 & 05 30 35.81  & 2.57 &  367.9 & 0.16 & 0.27  &  3.9e$-$03 & 0.34 &    5.49  &A0543    & 0.1754 &   \nl
 &$-$22 27 56.0 &      &        & 0.02 & $+$0.05 $-$0.05 &             & 0.13 &    1.65   &A  & (13)  &  \nl \nl
050 & 05 32 23.42  & 11.1 &  456.4 & 0.30 & 0.40  &  3.0e$-$02 & 0.63 &    6.35  &A0545    & 0.1540 &   \nl 
 &$-$11 32 03.5 &      &        & 0.03 & $+$0.05 $-$0.05 &             & 0.07 &    0.73   &A  & (4)  &  \nl \nl
051 & 05 33 14.02  & 2.93 &  505.8 & 0.15 & 0.27  &  3.9e$-$03 & 0.35 &    0.33  &S0535    & 0.0473 &   \nl 
 &$-$36 19 10.5 &      &        & 0.02 & $+$0.07 $-$0.06 &             & 0.08 &    0.08   &A  & (13)  &  \nl \nl
052 & 05 38 15.36  & 4.00 &  497.5 & 0.20 & 0.36  &  1.1e$-$03 & 0.47 &    1.69  &A3358    & 0.0915 & 12\nl 
 &$-$20 37 34.5 &      &        & 0.02 & $+$0.08 $-$0.06 &             & 0.09 &    0.33   &A  & (14)  &  \nl \nl
053 & 05 40 06.82  & 3.53 &  601.0 & 0.43 & 0.70  &  1.0e$-$04 & 0.92 &    0.50  &S0540    & 0.0358 &   \nl 
 &$-$40 50 30.5 &      &        & 0.03 & $+$0.08 $-$0.07 &             & 0.10 &    0.05   &A  & (2)  &  \nl \nl
054 & 05 47 37.44  & 1.95 &  579.3 & 0.31 & 0.47  &  1.2e$-$03 & 0.59 &    5.50  &A3364    & 0.1483 &   \nl 
 &$-$31 52 30.0 &      &        & 0.02 & $+$0.06 $-$0.05 &             & 0.07 &    0.66   &A  & (6) &  \nl \nl
055 & 05 48 36.70  & 1.88 &  553.8 & 0.26 & 0.75  &  3.9e$-$06 & 0.94 &    0.70  &A0548    & 0.0416 &   \nl
 &$-$25 28 27.0 &      &        & 0.02 & $+$0.26 $-$0.18 &             & 0.28 &    0.21   &A  & (1)  &  \nl \nl
056 & 05 52 52.08  & 4.33 &  529.0 & 0.33 & 0.53  &  5.3e$-$04 & 0.70 &   2.94  &A0550     & 0.0990 &   \nl 
 &$-$21 03 20.5 &      &        & 0.03 & $+$0.07 $-$0.06 &             & 0.09 &    0.37   &A  & (6) &  \nl \nl
057 & 05 57 11.89  & 3.95 &  705.1 & 0.21 & 0.38  &  2.3e$-$04 & 0.50 &  0.41  &S0555  & 0.0440 &   \nl 
 &$-$37 28 26.0 &      &        & 0.02 & $+$0.06 $-$0.05 &             & 0.08 &    0.06   &A  & (6) &  \nl \nl
058 & 06 00 27.27  & 5.68 &  993.4 & 0.27 & 0.29  &  4.1e$-$01 & 0.41 &         & \nodata  &\nodata & 13\nl 
 &$-$48 46 02.0 &      &        & 0.02 & $+$0.02 $-$0.02 &             & 0.03 &          &C  &      &  \nl \nl
059 & 06 01 37.77  & 5.01 &  749.3 & 0.26 & 1.10  &  1.0e$-$08 & 1.50 &    2.15  &A3376    & 0.0455 & 14\nl 
 &$-$40 00 31.0 &      &        & 0.02 & $+$0.38 $-$0.31 &             & 2.69 &    2.38   &A  & (2)  &  \nl \nl
060 & 06 05 52.68  & 4.30 &  714.7 & 0.38 & 0.46  &  8.7e$-$02 & 0.61 &    5.15  &A3378    & 0.1410 & 15\nl 
 &$-$35 18 08.0 &      &        & 0.02 & $+$0.04 $-$0.04 &             & 0.05 &    0.40   &A  & (15)  &  \nl \nl
061 & 06 21 44.17  & 5.17 & 1038.3 & 0.14 & 0.30  &  5.3e$-$05 & 0.41 &  0.40  &MS0620.6-5239   & 0.0480 &  \nl 
 &$-$52 42 12.0 &      &        & 0.01 & $+$0.07 $-$0.06 &             & 0.09 &    0.09   &A  & (10)  &  \nl \nl
062 & 06 22 16.78  & 5.57 & 3776.8 & 0.12 & 0.32  &  3.7e$-$11 & 0.44 &    0.11  &S0585    & 0.0241 & 16\nl 
 &$-$64 56 31.5 &      &        & 0.01 & $+$0.05 $-$0.04 &             & 0.06 &    0.02   &A  & (2)  &  \nl \nl
063 & 06 25 42.45  & 7.08 &  720.8 & 0.08 & 0.27  &  2.6e$-$03 & 0.39 &    0.19  &A3390    & 0.0338 & 17\nl
 &$-$37 15 02.0 &      &        & 0.01 &  $+$0.29 $-$0.11 &            & 0.31 &    0.15   &A  & (16)  &  \nl \nl
064 & 06 26 20.10  & 5.42 & 1185.9 & 0.39 & 0.95  &  1.5e$-$11 & 1.31 &    1.58  &A3391    & 0.0531 &   \nl 
 &$-$53 41 44.5 &      &        & 0.02 & $+$0.13 $-$0.11 &             & 0.17 &    0.20   &A  & (2)  &  \nl \nl
065 & 06 27 38.83  & 5.42 & 1327.9 & 0.25 & 0.81  &  8.3e$-$11 & 1.12 &    1.19  &A3395    & 0.0498 & 18\nl
 &$-$54 26 38.5 &      &        & 0.01 & $+$0.21 $-$0.15 &             & 0.25 &    0.27   &A  & (2)  &  \nl \nl
066 & 06 28 50.19  & 6.27 &  633.1 & 0.23 & 0.26  &  3.2e$-$01 & 0.37 &    4.81  &A3396    & 0.1759 &   \nl 
 &$-$41 43 32.5 &      &        & 0.02 & $+$0.03 $-$0.03 &             & 0.04 &    0.51   &A  & (6)  &  \nl \nl
067 & 06 38 46.66  & 6.57 &  861.7 & 0.28 & 0.36  &  2.7e$-$02 & 0.52 &   10.67  &S0592    & 0.2216 &   \nl 
 &$-$53 58 22.0 &      &        & 0.02 & $+$0.03 $-$0.03 &             & 0.05 &    0.92   &A  & (6)  &  \nl \nl
068 & 06 45 29.21  & 6.57 &  723.8 & 0.24 & 0.35  &  2.8e$-$03 & 0.49 &    5.82  &A3404    & 0.1670 &   \nl 
 &$-$54 13 17.5 &      &        & 0.02 & $+$0.04 $-$0.04 &             & 0.06 &    0.69   &A  & (6)  &  \nl \nl
069 & 06 58 30.41  & 6.34 &  507.8 & 0.35 & 0.51  &  2.2e$-$03 & 0.71 &   27.57  &1ES 0657-558 & 0.2994 & 19\nl 
 &$-$55 56 47.0 &      &        & 0.03 & $+$0.06 $-$0.06 &             & 0.09 &    3.18   &A  & (6)  &  \nl \nl
070 & 19 12 42.19  & 6.68 &  218.9 & 0.26 & 0.33  &  1.2e$-$01 & 0.47 &    1.09  &S0810    & 0.0736 &   \nl 
 &$-$75 17 24.0 &      &        & 0.04 & $+$0.06 $-$0.06 &             & 0.09 &    0.20   &A  & (6) &  \nl \nl
071 & 19 25 26.27  & 6.59 &  301.0 & 0.25 & 0.40  &  1.0e$-$02 & 0.56 &    1.43  &A3638    & 0.0774 &   \nl
 &$-$42 57 12.5 &      &        & 0.03 & $+$0.09 $-$0.08 &             & 0.12 &    0.30   &A  & (6) &  \nl \nl
072 & 19 52 09.89  & 4.86 &  191.1 & 0.20 & 0.43  &  1.3e$-$02 & 0.59 &    0.90  &A3651    & 0.0599 &   \nl
 &$-$55 03 18.5 &      &        & 0.04 & $+$0.22 $-$0.14 &             & 0.25 &    0.38   &A  & (1)  &  \nl \nl
073 & 20 12 35.08  & 4.59 &  174.9 & 0.91 & 2.44  &  1.6e$-$06 & 3.29 &    4.35  &A3667    & 0.0556 &   \nl 
 &$-$56 50 30.5 &      &        & 0.07 & $+$0.69 $-$0.50 &             & 0.81 &    1.07   &A  & (1)  &  \nl \nl
074 & 20 14 49.98  & 7.40 &  367.1 & 0.20 & 0.28  &  3.1e$-$02 & 0.41 &    4.08  & \nodata & 0.1538 &  20\nl 
 &$-$24 30 35.0 &      &        & 0.03 & $+$0.05 $-$0.05 &             & 0.07 &    0.71   &B  & (6) &  \nl \nl
075 & 20 18 41.52  & 4.72 &  346.3 & 0.29 & 0.55  &  7.1e$-$04 & 0.74 &    0.80  &S0861    & 0.0504 &   \nl 
 &$-$52 42 28.5 &      &        & 0.03 & $+$0.12 $-$0.10 &             & 0.15 &    0.16   &A  & (6) &  \nl \nl
076 & 20 22 59.09  & 5.59 &  179.6 & 0.16 & 0.25  &  7.1e$-$02 & 0.35 &    0.48  &S0868    & 0.0564 &  21\nl 
 &$-$20 57 25.0 &      &        & 0.04 & $+$0.10 $-$0.08 &             & 0.12 &    0.17   &A  & (6) &  \nl \nl
077 & 20 34 19.52  & 3.90 &  294.6 & 0.14 & 0.26  &  2.4e$-$02 & 0.35 &    1.24  &A3693    & 0.0910 &   \nl 
 &$-$34 29 12.5 &      &        & 0.03 & $+$0.11 $-$0.08 &             & 0.13 &    0.45   &A  & (1)  &  \nl \nl
078 & 20 34 41.41  & 3.90 &  256.6 & 0.33 & 0.48  &  1.9e$-$02 & 0.64 &    2.35  &A3694    & 0.0929 &   \nl
 &$-$34 04 16.0 &      &        & 0.04 & $+$0.09 $-$0.08 &             & 0.11 &    0.41   &A  & (17)  &  \nl \nl
079 & 20 34 46.86  & 3.56 &  309.9 & 0.42 & 0.95  &  4.4e$-$05 & 1.25 &    4.24  &A3695    & 0.0893 &   \nl 
 &$-$35 49 07.5 &      &        & 0.04 & $+$0.24 $-$0.19 &             & 0.28 &    0.96   &A  & (1)  &  \nl \nl
080 & 21 02 10.33  & 5.33 &  428.1 & 0.22 & 0.28  &  7.0e$-$02 & 0.38 &    5.70  &EXO2059-247 & 0.1880 &  \nl 
 &$-$24 31 59.5 &      &        & 0.02 & $+$0.04 $-$0.04 &             & 0.06 &    0.81   &A  & (18)  &  \nl \nl
081 & 21 04 19.81  & 3.58 &  354.8 & 0.21 & 0.31  &  1.7e$-$02 & 0.40 &    4.72  &A3739    & 0.1661 &   \nl 
 &$-$41 21 07.0 &      &        & 0.03 & $+$0.06 $-$0.06 &             & 0.08 &    0.89   &A  & (6) &  \nl \nl
082 & 21 04 51.57  & 3.08 &  278.4 & 0.37 & 0.66  &  1.1e$-$03 & 0.85 &    0.88  &ESO235-G050 & 0.0491 &  22\nl 
 &$-$51 49 21.0 &      &        & 0.04 & $+$0.14 $-$0.12 &             & 0.16 &    0.17   &B  & (19)  &  \nl \nl
083 & 21 07 07.88  & 5.52 &  427.9 & 0.13 & 0.30  &  3.5e$-$03 & 0.41 &    0.26  &A3744    & 0.0381 &   \nl 
 &$-$25 26 40.5 &      &        & 0.02 & $+$0.13 $-$0.09 &             & 0.15 &    0.10   &A  & (1)  &  \nl \nl
084 & 21 27 05.82  & 4.78 &  390.8 & 0.17 & 0.28  &  2.9e$-$03 & 0.38 &    6.11  &A2345    & 0.1760 &   \nl 
 &$-$12 10 12.5 &      &        & 0.02 & $+$0.05 $-$0.05 &             & 0.13 &    1.71   &A  & (20)  &  \nl \nl
085 & 21 29 40.08  & 4.22 &  286.0 & 0.28 & 0.35  &  8.8e$-$02 & 0.47 &         & \nodata  & \nodata &   23\nl
 &$-$00 05 49.5 &      &        & 0.03 & $+$0.05 $-$0.05 &             & 0.07 &           &C  &       &  \nl \nl
086 & 21 36 12.40  & 3.21 &  388.4 & 0.55 & 0.67  &  7.1e$-$02 & 0.87 &         & \nodata  & \nodata &   24\nl 
 &$-$62 22 24.5 &      &        & 0.04 & $+$0.06 $-$0.06 &             & 0.08 &          &C  &        &  \nl \nl
087 & 21 43 59.42  & 3.42 &  353.1 & 0.35 & 0.50  &  9.3e$-$03 & 0.66 &    1.86  &MRC2140-568 & 0.0815 &  \nl
 &$-$56 37 31.0 &      &        & 0.03 & $+$0.07 $-$0.07 &             & 0.09 &    0.25   &A  & (6) &  \nl \nl
088 & 21 45 54.95  & 4.03 &  258.8 & 0.20 & 0.38  &  8.4e$-$03 & 0.50 &    1.40  &A2377    & 0.0808 &   \nl 
 &$-$10 06 24.5 &      &        & 0.03 & $+$0.12 $-$0.09 &             & 0.14 &    0.40   &A  & (4)  &  \nl \nl
089 & 21 46 23.88  & 2.57 &  355.8 & 0.24 & 0.42  &  2.4e$-$03 & 0.54 &    1.35  &A3806    & 0.0765 &   \nl 
 &$-$57 17 22.5 &      &        & 0.03 & $+$0.10 $-$0.08 &             & 0.11 &    0.28   &A  & (1)  &  \nl \nl
090 & 21 46 55.7   & 1.77 &  312.7 & 0.27 & 0.49  &  1.6e$-$03 & 0.62 &    1.01  &A3809    & 0.0620 &   \nl 
 &$-$43 53 21   &      &        & 0.03 & $+$0.12 $-$0.10 &             & 0.14 &    0.22   &A  & (1)  &  \nl \nl
091 & 21 47 49.08  & 2.72 &  304.7 & 0.20 & 0.30  &  2.1e$-$02 & 0.39 &    0.59  &S0974    & 0.0596 &   \nl 
 &$-$46 00 02.0 &      &        & 0.03 & $+$0.07 $-$0.06 &             & 0.08 &    0.12   &A  & (2)  &  \nl \nl
092 & 21 49 07.24  & 2.31 &  370.1 & 0.26 & 0.32  &  9.4e$-$02 & 0.41 &    2.41  &A3814    & 0.1177 &   \nl 
 &$-$30 41 52.5 &      &        & 0.03 & $+$0.04 $-$0.04 &             & 0.06 &    0.33   &A  & (5)  &  \nl \nl
093 & 21 51 55.91  & 4.27 &  250.8 & 0.11 & 0.31  &  1.7e$-$02 & 0.41 &    0.74  &A2382    & 0.0648 &   \nl
 &$-$15 43 19.0 &      &        & 0.02 & $+$0.32 $-$0.14 &             & 0.33 &    0.60   &A  & (4)  &  \nl \nl
094 & 21 52 20.92  & 3.05 &  349.8 & 0.36 & 0.73  &  1.1e$-$04 & 0.94 &    3.58  &A2384    & 0.0943 &   \nl 
 &$-$19 33 54.5 &      &        & 0.03 & $+$0.16 $-$0.13 &             & 0.18 &    0.70   &A  & (4)  &  \nl \nl
095 & 21 54 10.21  & 2.12 &  362.7 & 0.44 & 0.87  &  3.0e$-$05 & 1.10 &    2.70  &A3822    & 0.0759 &   \nl
 &$-$57 52 05.5 &      &        & 0.04 & $+$0.16 $-$0.13 &             & 0.19 &    0.46   &A  & (1)  &  \nl \nl
096 & 21 58 23.58  & 2.77 &  375.2 & 0.19 & 0.50  &  5.3e$-$04 & 0.65 &    1.90  &A3825     & 0.0751 &   \nl 
 &$-$60 25 40.0 &      &        & 0.02 & $+$0.14 $-$0.12 &             & 0.42 &    1.02   &A  & (1)  &  \nl \nl
097 & 21 58 30.03  & 4.00 &  327.4 & 0.21 & 0.35  &  7.6e$-$03 & 0.46 &    1.28  &A2402    & 0.0806 &   \nl 
 &$-$09 47 54.5 &      &        & 0.03 &  $+$0.08 $-$0.07 &            & 0.10 &    0.28   &A  & (21)  &  \nl \nl
098 & 22 01 50.26  & 2.61 &  250.9 & 0.13 & 0.29  &  1.7e$-$02 & 0.37 &    0.77  &S0987    & 0.0701 &   \nl 
 &$-$22 25 53.5 &      &        & 0.03 & $+$0.18 $-$0.10 &             & 0.18 &    0.38   &A  & (5)  &  \nl \nl
099 & 22 01 58.85  & 2.84 &  372.0 & 0.69 & 1.18  &  2.9e$-$05 & 1.52 &    6.25  &A3827    & 0.0984 &   \nl 
 &$-$59 57 37.0 &      &        & 0.04 & $+$0.13 $-$0.12 &             & 0.17 &    0.69   &A  & (1)  &  \nl \nl
100  & 22 05 39.71  & 4.68 &  323.7 & 0.30 & 0.44  &  1.1e$-$02 & 0.60 &    0.91  &A2415    & 0.0597 &   \nl 
 &$-$05 35 00.5 &      &        & 0.03 & $+$0.07 $-$0.07 &             & 0.09 &    0.14   &A  & (4)  &  \nl \nl
101 & 22 09 25.52  & 2.06 &  348.8 & 0.17 & 0.31  &  6.4e$-$03 & 0.39 &    1.90  &A3836    & 0.1065 &   \nl
 &$-$51 50 37.5 &      &        & 0.02 & $+$0.09 $-$0.07 &             & 0.11 &    0.51   &A  & (6) &  \nl \nl
102 & 22 10 20.09  & 3.87 &  286.7 & 0.45 & 0.89  &  1.1e$-$04 & 1.17 &    3.51  &A2420    & 0.0838 &   \nl 
 &$-$12 10 49.0 &      &        & 0.04 & $+$0.18 $-$0.15 &             & 0.22 &    0.66   &A  & (4)  &  \nl \nl
103 & 22 14 32.15  & 3.86 &  288.7 & 0.47 & 0.68  &  5.4e$-$03 & 0.90 &    3.66  &A2426    & 0.0978 &   \nl
 &$-$10 22 23.0 &      &        & 0.04 & $+$0.09 $-$0.09 &             & 0.12 &    0.47   &A  & (1)  &  \nl \nl
104 & 22 16 16.15  & 4.50 &  288.4 & 0.33 & 0.40  &  1.2e$-$01 & 0.54 &    1.65  &A2428    & 0.0846 &   \nl 
 &$-$09 20 11.0 &      &        & 0.04 & $+$0.05 $-$0.05 &             & 0.07 &    0.22   &A  & (22)  &  \nl \nl
105 & 22 16 56.23  & 2.28 &  283.0 & 0.24 & 0.31  &  6.6e$-$02 & 0.40 &    2.85  &\nodata   & 0.1301 &   \nl 
 &$-$17 25 25.5 &      &        & 0.03 & $+$0.06 $-$0.05 &             & 0.07 &    0.49   &A  & (8)   &  \nl \nl
106 & 22 17 45.58  & 1.10 &  333.6 & 0.25 & 0.34  &  3.9e$-$02 & 0.42 &    3.84  &A3854    & 0.1474 &   \nl 
 &$-$35 43 21.0 &      &        & 0.03 & $+$0.05 $-$0.05 &             & 0.06 &    0.59   &A  & (5)  &  \nl \nl
107 & 22 18 07.76  & 2.83 &  446.7 & 0.28 & 0.43  &  3.2e$-$03 & 0.55 &   2.11  &\nodata  & 0.0951 &  25\nl 
 &$-$65 12 00.0 &      &        & 0.03 & $+$0.07 $-$0.06 &             & 0.08 &    0.31   &A  & (6) &  \nl \nl
108 & 22 18 12.49  & 5.73 &  253.6 & 0.18 & 0.38  &  8.2e$-$03 & 0.52 &  1.80  &MS2215.7-0404 & 0.0900 &  26\nl 
 &$-$03 47 59.0 &      &        & 0.03 & $+$0.17 $-$0.11 &             & 0.20 &    0.69   &A  & (10)  &  \nl \nl
109 & 22 18 40.71  & 1.33 &  330.9 & 0.28 & 0.44  &  4.7e$-$03 & 0.54 &    3.67  &A3856    & 0.1260 &   \nl
 &$-$38 53 46.5 &      &        & 0.03 & $+$0.08 $-$0.07 &             & 0.09 &    0.63   &A  & (23)  &  \nl \nl
110 & 22 20 33.87  & 1.09 &  327.6 & 0.43 & 0.55  &  4.9e$-$02 & 0.67 &    6.77  &A3866    & 0.1544 &  27\nl 
 &$-$35 09 52.0 &      &        & 0.04 & $+$0.06 $-$0.06 &             & 0.07 &    0.75   &B  & (6) &  \nl \nl
111 & 22 23 48.91  & 5.34 &  237.9 & 0.29 & 0.63  &  1.4e$-$03 & 0.87 &    3.03  &A2440    & 0.0904 &   \nl
 &$-$01 39 25.0 &      &        & 0.04 & $+$0.21 $-$0.15 &             & 0.25 &    0.89   &A  & (4)  &  \nl \nl
112 & 22 24 31.31  & 3.54 &  362.3 & 0.15 & 0.28  &  8.8e$-$03 & 0.36 & 0.95  &APM222041.3-552 & 0.0780 &   \nl 
 &$-$55 15 15.0 &      &        & 0.02 & $+$0.09 $-$0.07 &             & 0.10 &    0.27   &A  & (24)  &  \nl \nl
113 & 22 27 52.46  & 1.09 &  312.6 & 0.42 & 0.53  &  5.7e$-$02 & 0.65 &    0.90  &A3880    & 0.0570 &   \nl 
 &$-$30 34 10.5 &      &        & 0.04 & $+$0.06 $-$0.06 &             & 0.08 &    0.10   &A  & (23)  &  \nl \nl
114 & 22 28 55.32  & 2.22 &  458.9 & 0.13 & 0.27  &  4.6e$-$03 & 0.34 &    0.24  &ESO146-G028 & 0.0412 &  \nl 
 &$-$60 54 24.5 &      &        & 0.02 & $+$0.10 $-$0.07 &             & 0.11 &    0.08   &A  & (25)  &  \nl \nl
115 & 22 34 28.50  & 1.20 &  258.2 & 0.48 & 0.69  &  8.0e$-$03 & 0.85 &    8.20  &A3888    & 0.1510 &   \nl 
 &$-$37 43 53.5 &      &        & 0.04 & $+$0.09 $-$0.09 &             & 0.11 &    1.08   &A  & (26)  &  \nl \nl
116 & 22 35 39.73  & 5.81 &  166.6 & 0.24 & 0.36  &  4.0e$-$02 & 0.51 &    0.77  &A2457    & 0.0597 &   \nl 
 &$-$01 29 02.5 &      &        & 0.04 & $+$0.10 $-$0.09 &             & 0.13 &    0.20   &A  & (4)  &  \nl \nl
117 & 22 46 17.79  & 1.52 &  411.1 & 0.31 & 0.48  &  2.1e$-$03 & 0.60 &    2.43  &A3911    & 0.0974 &   \nl
 &$-$52 43 48.0 &      &        & 0.03 & $+$0.07 $-$0.07 &             & 0.09 &    0.35   &A  & (27)  &  \nl \nl
118 & 22 48 44.45  & 1.79 & 250.2 & 0.38 & 0.52 & 2.8e$-$02 & 0.65 & 17.18 &S1063 & 0.2520\tablenotemark{a}&\nl 
 &$-$44 31 50.0 &  &      & 0.04 & $+$0.08 $-$0.07 &             & 0.09 &    2.49   &B  & (2)        &  \nl \nl
119 & 22 50 03.61  & 2.80 &  469.6 & 0.42 & 0.93  &  1.1e$-$06 & 1.20 &    4.69  &A3921    & 0.0936 &   \nl
 &$-$64 26 30.0 &      &        & 0.03 & $+$0.20 $-$0.18 &             & 0.23 &    0.87   &A  & (1)  &  \nl \nl
120 & 22 54 01.45  & 2.23 &  400.2 & 0.19 & 0.28  &  1.6e$-$02 & 0.35 &    6.62  &AM2250-633 & 0.2112 &  \nl 
 &$-$63 15 02.0 &      &        & 0.02 & $+$0.06 $-$0.05 &             & 0.07 &    1.24   &A  & (6)  &  \nl \nl
121 & 22 54 27.79  & 2.12 &  404.7 & 0.11 & 0.27  &  7.3e$-$03 & 0.35 &          & \nodata &\nodata   &  28\nl 
 &$-$58 06 41.5 &      &        & 0.02 & $+$0.17 $-$0.09 &             & 0.17 &          &C  &       &  \nl \nl
122 & 23 13 58.42  & 1.85 &  202.4 & 0.87 & 1.07  &  7.0e$-$02 & 1.34 &    1.93  &S1101    & 0.0580 &   \nl 
 &$-$42 43 47.0 &      &        & 0.07 & $+$0.10 $-$0.10 &             & 0.13 &    0.18   &A  & (10)  &  \nl \nl
123 & 23 15 42.49  & 4.18 &  343.1 & 0.19 & 0.30  &  1.2e$-$02 & 0.40 &    0.12  &ZWIII99  & 0.0267 &  29\nl 
 &$-$02 22 27.0 &      &      & 0.03 & $+$0.07 $-$0.06 &             & 0.08 &    0.03   &A  & (8)    &  \nl \nl
124 & 23 21 36.11  & 1.96 &  164.7 & 0.29 & 0.52  & 1.1e$-$02 & 0.66 &    2.23  &A3998     & 0.0890 &   \nl 
 &$-$41 53 37.0 &      &        & 0.04 & $+$0.16 $-$0.13 &             & 0.18 &    0.62   &A  & (10)  &  \nl \nl
125 & 23 25 18.95  & 2.50 &  332.3 & 0.87 & 1.06  &  5.6e$-$02 & 1.36 &    4.21  &A2597    & 0.0852 &  30\nl
 &$-$12 07 32.0 &      &        & 0.05 & $+$0.08 $-$0.08 &             & 0.10 &    0.31   &A  & (4)  &  \nl \nl
126 & 23 44 15.98  & 3.54 &  339.8 & 0.40 & 0.93  &  1.8e$-$05 & 1.21 &    3.20  & \nodata  & 0.0786 &   \nl 
 &$-$04 22 24.5 &      &        & 0.04 & $+$0.23 $-$0.18 &             & 0.27 &    0.71   &A  & (8)   &  \nl \nl
127 & 23 47 41.78  & 1.55 &  334.1 & 1.30 & 2.21  &  2.0e$-$06 & 2.75 &    1.01  &A4038    & 0.0292 &   \nl
 &$-$28 08 26.5 &      &        & 0.06 & $+$0.19 $-$0.18 &             & 0.23 &    0.08   &A  & (1)  &  \nl \nl
128 & 23 51 40.36  & 1.66 &  329.8 & 0.43 & 0.52  &  8.8e$-$02 & 0.66 &   14.10  &A2667    & 0.2264 &  31\nl 
 &$-$26 04 54.0 &      &        & 0.04 & $+$0.06 $-$0.06 &             & 0.07 &    1.51   &A  & (8)   &  \nl \nl
129 & 23 54 12.70  & 2.92 &  317.9 & 0.34 & 0.60  &  7.8e$-$04 & 0.78 &    1.92  &A2670    & 0.0762 &   \nl 
 &$-$10 24 57.0 &      &        & 0.03 & $+$0.11 $-$0.10 &             & 0.14 &    0.34   &A  & (1)  &  \nl \nl
130 & 23 57 00.02  & 1.10 &  350.6 & 0.96 & 1.61  &  1.3e$-$05 & 1.97 &    1.79  &A4059    & 0.0460 &   \nl 
 &$-$34 45 24.5 &      &        & 0.05 & $+$0.15 $-$0.15 &             & 0.18 &    0.17   &A  & (10)  &  
\tablenotetext{a}{This is an estimated redshift.}
\tablecomments{Notes on single sources:
{\bf 1.} -- cluster in Strubble \& Rood (1991);
{\bf 2.} -- identified as galaxy MS0122.9+0129 in Maccacaro et al. (1994), extended 
    from ROSAT HRI and PSPC pointed observations, many small galaxies nearby, z measured
    from 3 galaxies; 
{\bf 3.} -- central galaxy is radio source (Brinkman, Siebert \& Boller 1994), 
    z measured from 3 galaxies;
{\bf 4.} -- X-ray emission slightly elongated towards cluster A3016; 
{\bf 5.} -- source at low z (measured on 2 galaxies) but pointlike,
    several other objects in vicinity, could be a group or a single galaxy, 
    uncertain identification;
{\bf 6.} -- extended in HRI observation, z measured from 2 galaxies, 
    RASS data slightly elongated towards a pointlike
    source at 02h32m37.5s -44d21m51s (J2000);
{\bf 7.} -- double peaked X-ray emission, SRT count rate underestimated because the RASS ML
    position is centered on one peak, $0.61\pm 0.01$ cts/s hard count rate from PSPC pointed observations;
{\bf 8.} -- extended in HRI observation, identified as normal galaxy in Einstein Catalog
    IPC Slew Survey but is cluster (4 galaxies with consistent z);
{\bf 9.} -- extended in HRI observation, paired with IC366 at 3.2$^\prime$, flagged
    as nearby galaxy group in Lyon-Meudon catalog (Garcia 1993) on the basis of a percolation method, 
    z available only for central galaxy NGC1550;
{\bf 10.} -- normal galaxy NGC1650 in rich field, likely galaxy group, only one z available;
{\bf 11.} -- z measured from 4 galaxies;
{\bf 12.} -- cluster behind star HD37493 (spectral type K0, m$_V$ = 8.2), which could only slightly 
     contribute to the X-ray emission (see Maccacaro et al. 1988), the extended X-ray emission implies 
     that most of the emission comes from the cluster;
{\bf 13.} -- unconfirmed cluster with no spectroscopic information available;
{\bf 14.} -- elongated X-ray emission between 2 bright galaxies, 
     SRT count rate in agreement with hard count rate from PSPC pointed observation;
{\bf 15.} -- extended in HRI observation;
{\bf 16.} -- also A3389 (optical position at 3.3$^\prime$ from X-ray peak), double peaked 
     X-ray emission centered on NGC 2235 (at 0.7$^\prime$); 
{\bf 17.} -- double peaked X-ray emission, RASS ML position centered on one peak, $0.27\pm 0.01$ cts/s
     hard count rate from PSPC pointed observation;
{\bf 18.} -- double peaked X-ray emission, RASS ML position centered on one peak, $1.16\pm 0.03$ cts/s
     hard count rate from PSPC pointed observation;
{\bf 19.} -- extended in HRI observation, classified as galaxy cluster in Tucker, Tananbaum 
     \& Remillard (1995);
{\bf 20.} -- probable cluster with only one z measured; 
{\bf 21.} -- central galaxy is radio source (Brinkman, Siebert
 \& Boller 1994, Douglas et al. 1996), 
     z measured from 2 galaxies;
{\bf 22.} -- extended in HRI observation, probable group, z only from the brightest galaxy;
{\bf 23.} -- extended in HRI observation, no spectroscopic information available;
{\bf 24.} -- unidentified candidate with no spectroscopic information available;
{\bf 25.} -- z measured from 5 galaxies;
{\bf 26.} -- MS2215.70404 (coincident with RASS ML emission peak) and MS2216.00401 are two separate 
     sources probably parts of the extended X-ray emission from a single cluster (Gioia \& Luppino 1994);
{\bf 27.} -- extended in HRI observation, only one z measured;
{\bf 28.} -- probable cluster, no spectroscopic information available;
{\bf 29.} -- extended in HRI observation, paired with NGC7556, z measured from 2 galaxies;
{\bf 30.} -- extended in HRI observation;
{\bf 31.} -- extended in HRI observation.
}
\tablerefs{
(1) Mazure et al. 1996;
(2) Abell et al. 1989;
(3) Dalton et al. (1997); 
(4) Strubble \& Rood 1987;
(5) Collins et al. 1995; 
(6) data from the ESOKP collaboration; 
(7) Zabludoff et al. 1993;
(8) data from the South Galactic Pole survey (see Romer et al. 1994);
(9) Crawford et al. 1995;
(10) Stocke et al. 1991;
(11) de Vaucouleurs et al. 1991;
(12) Huchra et al. 1993;
(13) Quintana \& Ramirez 1995;
(14) Henry \& Mullir 1997, priv. comm.;
(15) Ebeling et al. 1996;
(16) Galli et al. 1993;
(17) Muriel, Nicotra \& Lambas 1990;
(18) NED11 1992;
(19) Lauberts \& Valentijn 1989;
(20) Ledlow \& Owen 1995; 
(21) Postman, Huchra \& Geller 1992; 
(22) Allen et al. 1992;
(23) Ebeling \& Maddox 1995; 
(24) Dalton et al. 1994;
(25) Muriel, Nicotra \& Lambas 1995;
(26) Teague et al. 1990;
(27) Postman \& Lauer 1995.
}
\enddata
\end{deluxetable}
\clearpage

\begin{deluxetable}{lrrrrcrrrrr}
\tablewidth{40pc}
\tablecaption{Rejected Cluster Candidates}
\tablehead{
\colhead{Source~ ID} & \colhead{RA}       & \colhead{Dec}     & 
\colhead{F$_{X}$}        & \colhead{F$_{X}$ Err}    & \colhead{Type} \nl
\colhead{}           & \colhead{(J2000)}  & \colhead{(J2000)} & 
\colhead{10$^{-11}$ cgs} & \colhead{10$^{-11}$ cgs} & \colhead{ }}
\startdata
 HD3237           &  00 35 08.17 & $-$50 20 05.5 & 0.61 & 0.08 & STA \nl 
 NGC0253          &  00 47 32.70 & $-$25 17 21.0 & 0.45 & 0.07 & GAL \nl
 G60              &  01 35 00.88 & $-$29 54 40.0 & 1.54 & 0.06 & STA \nl
                  &  02 27 16.47 & $+$02 01 55.5 & 1.10 & 0.08 & AGN \nl
 2A0311$-$227     &  03 14 12.88 & $-$22 35 41.5 & 1.81 & 0.11 & STA \nl
 MRK0609          &  03 25 26.58 & $-$06 08 20.5 & 0.45 & 0.04 & AGN \nl
 HR1325           &  04 15 22.25 & $-$07 38 56.5 & 0.51 & 0.07 & STA \nl
 H0449$-$55       &  04 53 31.51 & $-$55 52 01.0 & 0.79 & 0.05 & STA \nl
 HR702            &  05 12 55.25 & $-$16 12 19.5 & 0.35 & 0.06 & STA \nl
 NGC1851          &  05 14 06.18 & $-$40 02 31.0 & 2.87 & 0.10 & GC  \nl
 0548$-$322       &  05 50 40.53 & $-$32 16 17.5 & 2.04 & 0.08 & AGN \nl
 HD45081          &  06 18 27.59 & $-$72 02 40.5 & 0.74 & 0.04 & STA \nl
 HR2326           &  06 23 57.31 & $-$52 41 43.0 & 0.50 & 0.04 & STA \nl
 PMNJ1931$-$2635  &  19 31 50.01 & $-$26 34 31.5 & 0.45 & 0.09 & AGN \nl
 PKS1930$-$510    &  19 34 51.95 & $-$50 52 54.0 & 0.50 & 0.54 & GAL \nl
 HR7571           &  19 53 06.63 & $-$14 36 08.0 & 0.44 & 0.06 & STA \nl
 PKS2005$-$489    &  20 09 25.07 & $-$48 49 48.0 & 1.62 & 0.09 & AGN \nl
 PKS2035$-$714    &  20 40 06.08 & $-$71 14 53.0 & 0.35 & 0.06 & AGN \nl
                  &  20 41 49.71 & $-$37 33 45.0 & 0.57 & 0.10 & GAL
\tablenotemark{a} \nl
 HD205249         &  21 34 16.18 & $-$13 29 07.5 & 0.54 & 0.07 & STA \nl
 PMNJ2150$-$1411  &  21 50 15.62 & $-$14 10 45.0 & 0.85 & 0.10 & AGN \nl
 ESO075$-$G041    &  21 56 54.04 & $-$69 40 34.5 & 0.38 & 0.06 & AGN \nl
 SAO145804        &  22 00 36.51 & $-$02 44 27.0 & 5.86 & 0.22 & STA \nl
 [HB89]2227$-$399 &  22 30 39.28 & $-$39 42 54.0 & 0.39 & 0.05 & AGN \nl
 NGC7603          &  23 18 56.18 & $+$00 14 38.5 & 0.40 & 0.04 & AGN \nl
 HD220054         &  23 21 52.89 & $-$69 42 18.0 & 0.40 & 0.07 & STA \nl
 HD220186         &  23 21 55.46 & $-$10 50 03.0 & 0.56 & 0.06 & STA \nl
\tablecomments{Units of right ascension are hours, minutes, and seconds,
and units of declination are degrees, arcminutes, and, arcseconds. See
text for explanation of types.}
\tablenotetext{a}{Identification based on ROSAT HRI and COSMOS data.}
\enddata
\end{deluxetable}
\clearpage

\begin{deluxetable}{lrrccccc}
\tablewidth{30pc}
\tablecaption{``Missed'' ACO Clusters}
\tablehead{
\colhead{Name}&\colhead{RA(J2000)}&\colhead{Dec(J2000)}&\colhead{$cr_{SASS1}$}& 
z & Ref. \nl
\colhead{ } &\colhead{[deg]} &\colhead{[deg]} & [.1-2.4 keV] & & }
\startdata
A0194 &  21.3867&  -1.5069& 0.045   & 0.0178 & (1) \nl
A0514 &  71.9158& -20.4290& 0.026   & 0.0730 & (1) \nl
A3164 &  56.4567& -57.0456& \nodata & 0.0611 & (2) \nl
A3223 &  62.1429& -30.8189& 0.036   & 0.0601 & (3) \nl
A3716 & 312.8858& -52.7122& 0.051   & 0.0456 & (2) \nl
A3733 & 315.4387& -28.0283& 0.034   & 0.0386 & ~~(2)
\tablerefs{
(1) Struble \& Rood 1987; 
(2) Abell et al. 1989; 
(3) den Hartog \& Katgert 1996.}
\enddata
\end{deluxetable}

\end{document}